\documentclass[aip, jap, reprint]{revtex4-2}

\usepackage[utf8]{inputenc}
\usepackage[T1]{fontenc}
\usepackage[english]{babel}

\usepackage{hyperref}
\usepackage{xcolor}
	\definecolor{vert}{rgb}{0.09,0.59,0.18}
	\colorlet{bluyan}{blue!60!cyan}

\usepackage{amsmath}
\usepackage{amssymb}
\usepackage[b]{esvect}
\usepackage{siunitx}
	\DeclareSIUnit{\kth}{\watt\per\meter\per\kelvin}		
	\DeclareSIUnit{\cpMol}{\joule\per\mole\per\kelvin}	
	\DeclareSIUnit{\sig}{\siemens\per\meter}				
	\DeclareSIUnit{\degC}{\degreeCelsius}				
\usepackage[version=4]{mhchem}

\usepackage{graphicx}
\usepackage[caption=false]{subfig}		
\usepackage{tikz}
	\usetikzlibrary{decorations.pathreplacing, calligraphy}
\usepackage{tikz-3dplot}
\usepackage{booktabs}

\newcommand{\kth}{k_{th}}
\newcommand{\ra}{\rightarrow}
\newcommand{\fpartial}[2]{\frac{\partial #1}{\partial #2}}
\newcommand{\vbl}{\vspace{\baselineskip}}	

\begin{document}
\title{Multi-Physics Modeling Of Phase Change Memory Operations in Ge-rich \ce{Ge2Sb2Te5} Alloys}

\author{Robin Miquel}
\affiliation{STMicroelectronics, 850 rue Jean Monnet, 38926 Crolles, France}
\affiliation{Univ. Grenoble Alpes, CEA, Leti, F-38000 Grenoble, France}
\affiliation{Laboratoire de Physique de la Matière Condensée, Ecole Polytechnique, CNRS, Institut Polytechnique de Paris, 91128 Palaiseau, France}

\author{Thomas Cabout}
\affiliation{STMicroelectronics, 850 rue Jean Monnet, 38926 Crolles, France}

\author{Olga Cueto}
\affiliation{Univ. Grenoble Alpes, CEA, Leti, F-38000 Grenoble, France}

\author{Benoit Sklénard}
\affiliation{Univ. Grenoble Alpes, CEA, Leti, F-38000 Grenoble, France}

\author{Mathis Plapp}
\affiliation{Laboratoire de Physique de la Matière Condensée, Ecole Polytechnique, CNRS, Institut Polytechnique de Paris, 91128 Palaiseau, France}

\begin{abstract}
One of the most widely used active materials for phase-change memories (PCM), the ternary 
stoichiometric compound \ce{Ge2Sb2Te5} (GST), has a low crystallization temperature of
around \qty{150}{\degC}. One solution to achieve higher operating temperatures is to enrich
GST with additional germanium (GGST). This alloy crystallizes into a polycrystalline mixture of
two phases, GST and almost pure germanium. 
In a previous work [R. Bayle {\it et al.}, J. Appl. Phys. {\bf 128}, 185101 (2020)], 
this crystallization process was studied using a multi-phase 
field model (MPFM) with a simplified thermal field calculated by a separate solver. Here, we 
combine the MPFM and a phase-aware electro-thermal solver to achieve
a consistent multi-physics model for device operations in PCM.
Simulations of memory operations 
are performed to demonstrate its ability to reproduce experimental observations and
the most important calibration curves that are used to assess the performance of a PCM cell.
\end{abstract}

\maketitle

\section{Introduction}
Phase change memories (PCM) relying on the electrical contrast between amorphous and 
crystalline phases of chalcogenide materials have been identified as a promising 
solution for embedded non-volatile memory technologies \cite{redaelli2022}.
In such devices, the state of individual memory cells is controlled using short but
intense pulses of electric current. The Joule heating generated by these pulses
can locally melt the material, and the shape of the pulse controls the cooling rate.
The latter determines whether the material ends up in a crystalline or amorphous state.

Materials for PCM must therefore fulfill several requirements: they must have a strong
contrast of resistivity between the amorphous and the crystalline state, but at the
same time they must have a rapid crystallization kinetics to allow for fast switching 
of memory states. The stoichiometric compound \ce{Ge2Sb2Te5} (GST), originally 
developed for rewritable optical disks \cite{yamada1991}, is the most 
widely used active material. 
However, for some applications (namely, for the automotive market) memories must withstand high temperatures (\qty{150}{\degC}) for several years.
GST is not compatible with such requirements because its crystallization temperature, the temperature at which the amorphous phase spontaneously crystallizes, is relatively low (around \qty{150}{\degC}).
In case of a spontaneous and unintentional crystallization, the stored information is lost.

One possibility to increase the crystallization temperature is to enrich GST with
additional germanium \cite{zuliani2013}. While this Ge-rich GST (GGST) allies 
excellent crystallization temperature with a reasonably high switching speed, 
it exhibits germanium segregation \cite{luong2021},
which leads to the nucleation of a new crystalline phase, almost pure germanium.
The active material becomes a two-phase polycrystal, and this complex structure
can have a large impact on device performance. Predictive modeling of phase 
change and segregation would thus be of great benefit to aid technological development 
and to improve the understanding of the underlying mechanisms.

In a previous work, a multi-phase field model (MPFM) has been developed to simulate 
the evolution of GGST microstructure \cite{bayle2020}. In this formalism, the local
presence of each phase is indicated by a scalar phase field, and the grain structure is
described by a field that indicates the local crystal orientation. The evolution
of all the fields is coupled to the diffusion of chemical elements, which
is driven by gradients of appropriate chemical potentials. The latter depend on
temperature, such that the time-dependent temperature field is required for
a realistic simulation of memory operations. In our previous work~\cite{bayle2020}, 
the temperature field for a given electric pulse shape was obtained by a separate 
electrothermal solver. This weak (one-way) coupling neglects the influence of 
the phase distribution on the electric current, which induces inhomogeneous
heating. This precludes a quantitative modeling of memory operations.

In this article, we present a more complete model for phase change memories, obtained 
by a full coupling of the MPFM with an electro-thermal model. More precisely, the
equations for electrical and heat conduction are added to the MPFM model, with
phase-dependent transport coefficients. This model can account for 
the multi-physics nature of memory operations and the influence of the material 
structure on the performances of the memory cell. Since the coupled system of
equations is numerically stiff, numerical methods with multiple grids and 
timesteps are required for efficient simulations. 

The model contains a large number of materials parameters, which are gathered
from the literature or taken from in-house measurements. Comparison of the
simulations with measurements performed on memory devices allows us to assess
the quality of the model and to improve on several parametrizations.
Notably, it turns out that the thermal boundary resistances (TBR) at the interfaces between different materials have to be taken into account. TBR lead to discontinuities in
temperature between the two sides of an interface when the interface is crossed
by a heat flux. In macroscopic systems, TBR are usually negligible, but in
PCM devices with characteristic size of a few tens of nanometers and high heat 
fluxes, considerable temperature jumps can occur \cite{durai2020}.
The experimental device characteristics can only be reproduced by our model if TBR 
are included both on the interfaces between the different parts of the device 
and on the internal interfaces between different phases of the active material.
This is one of the main new results of our work.

Some preliminary aspects of this work have already been published \cite{miquel2023};
here, the full details of the model as well as new results are presented. The paper
starts by the description of the simulation setup and the model equations in
Sec.~\ref{sec_model}. The equations contain numerous parameters, and 
Sec.~\ref{sec_param} discusses our choices, taking into account the
existing literature. Some details concerning the numerical implementation
are discussed in Sec.~\ref{sec_numerics}, and the main results are presented
in Sec.~\ref{sec_results}.

\section{Models}
\label{sec_model}
\subsection{Geometry and simulation setup}

\begin{figure}[!b]
	\centering
	\tdplotsetmaincoords{82}{115} 
	\begin{tikzpicture}
		[tdplot_main_coords,
			TE/.style={opacity=1,fill=orange!70},
			PCM/.style={fill=blue!50},
			Heater/.style={fill=red!80},
			Oxyde/.style={fill=yellow!60},
			Front/.style={canvas is yz plane at x=\lx, transform shape}]


		\def\lx{3.5};		
		\def\ly{6};		
		\def\lyH{.15};	
		\def\lzTE{.82};	
		\def\lzPC{1.1};	
		\def\rZA{.85};	
		\def\lzH{1.8};	
		\def\lzO{1.55};	
		\def\lzBE{.42};	

		\def\zTE{\lzPC};			
		\def\zzTE{\lzPC + \lzTE};	
		\def\yH{\ly*0.5 - \lyH*0.5};	
		\def\yyH{\ly*0.5 + \lyH*0.5};	
		\def\zBE{-\lzH - \lzBE};	
		\def\zzBE{-\lzH};			
		\def\yZA{\ly*0.5 - \rZA}	
		\def\yyZA{\ly*0.5 + \rZA}	

		\draw[Heater] (\lx,\yyH,-\lzO) -- (\lx,\ly,-\lzO) -- (\lx,\ly,-\lzH)
		           -- (\lx,\yH,-\lzH) -- (\lx,\yH,0) -- (\lx,\yyH,0) -- cycle;
		\draw[Heater] (\lx,\yyH,-\lzO) -- (\lx,\yyH,0) -- (0,\yyH,0) -- (0,\yyH,-\lzO) -- cycle;
		\draw[Heater] (\lx,\yyH,-\lzO) -- (0,\yyH,-\lzO) -- (0,\ly,-\lzO) -- (\lx,\ly,-\lzO) -- cycle;
		\draw[Heater] (\lx,\ly,-\lzO) -- (0,\ly,-\lzO) -- (0,\ly,-\lzH) -- (\lx,\ly,-\lzH) -- cycle;

		\draw[Oxyde] (\lx,0,0) -- (\lx,\yH,0) -- (\lx,\yH,-\lzH) -- (\lx,0,-\lzH) -- cycle;

		\draw[Oxyde] (0,\yyH,0) -- (0,\ly,0) -- (0,\ly,-\lzO) -- (0,\yyH,-\lzO) -- cycle;
		\draw[Oxyde, opacity=0.7] (0,\ly,0) -- (0,\ly,-\lzO) -- (\lx,\ly,-\lzO) -- (\lx,\ly, 0) -- cycle;
		\draw[Oxyde, opacity=0.7] (\lx,\yyH,0) -- (\lx,\ly,0) -- (\lx,\ly,-\lzO) -- (\lx,\yyH,-\lzO) -- cycle;

		\draw[TE] (\lx,\ly,\zzTE) -- (\lx,\ly,\zTE) -- (\lx,0,\zTE) -- (\lx,0,\zzTE) -- cycle;
		\draw[TE] (\lx,\ly,\zzTE) -- (\lx,0,\zzTE) -- (0,0,\zzTE) -- (0,\ly,\zzTE) -- cycle;
		\draw[TE] (\lx,\ly,\zzTE) -- (0,\ly,\zzTE) -- (0,\ly,\zTE) -- (\lx,\ly,\zTE) -- cycle;

		\draw[PCM] (\lx,\ly,\lzPC) -- (\lx,\ly,0) -- (\lx,0,0) -- (\lx,0,\lzPC) -- cycle;
		\draw[PCM] (\lx,\ly,\lzPC) -- (0,\ly,\lzPC) -- (0,\ly,0) -- (\lx,\ly,0) -- cycle;

		\def\toRad{0.00278};
		\draw [Front, fill = blue!20, dashed]
			(\yyZA,0) to [out=90, in=0] (\ly*0.5, \rZA) to [out=180, in=90] (\yZA,0);
		\draw [Front]
			(\yyZA,0) -- (\yZA,0);

		\draw[TE] (\lx,\ly,\zzBE) -- (\lx,\ly,\zBE) -- (\lx,0,\zBE) -- (\lx,0,\zzBE) -- cycle;
		\draw[TE] (\lx,\ly,\zzBE) -- (0,\ly,\zzBE) -- (0,\ly,\zBE) -- (\lx,\ly,\zBE) -- cycle;

		\draw[Front, ->, very thick] (1.52, -\lzH*0.45) -- (\yH - 0.15, -\lzH*0.5);
		\node[anchor=east] at (\lx, 1.45, -\lzH*0.41){Heater};
		\draw[Front, ->, very thick] (\ly*.7, \zzTE*2) -- (\ly*.53, \lzPC*.5);
		\node at (\lx, \ly*.7, \zzTE*2.37){Active volume};
	\end{tikzpicture}
	\caption[Structure Wall d'une PCM]
	{Schematic view of the Wall structure. The top and bottom electrodes are drawn in orange,
	the heater in red, the oxides in yellow, the phase-change material in blue, and the active
	zone in light blue. Drawing adapted from figure 3 in [\onlinecite{sousa2015}].}
	\label{fig_structure_wall}
\end{figure}
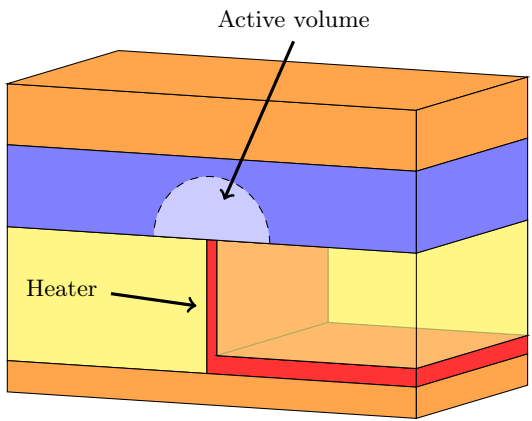

\begin{figure}[!ht]
	\centering
	\begin{tikzpicture}
		\def\scale{0.0193}		
		\def\lx{   300 * \scale};	
		\def\lxH{    5 * \scale};	
		\def\lxAc{ 150 * \scale};	
		\def\lyTE{ 100 * \scale};	
		\def\lyPCM{ 50 * \scale};	
		\def\lyMH{  75 * \scale};	
		\def\lyBE{  18 * \scale};	
		\pgfmathsetmacro\ly{\lyTE + \lyPCM + \lyMH + \lyBE}	
		\pgfmathsetmacro\lxM{(\lx - \lxH) / 2}		
		\pgfmathsetmacro\lxNAc{(\lx - \lxAc) / 2}	
		\pgfmathsetmacro\xHD{\lxM + \lxH}			
		\pgfmathsetmacro\yTE{\ly - \lyTE}			
		\pgfmathsetmacro\yPCM{\lyMH + \lyBE}		

		\draw (0, 0) rectangle (\lx, \ly);
		\draw (0, \yTE)  -- (\lx, \yTE);	
		\draw (0, \yPCM) -- (\lx, \yPCM);	
		\draw (0, \lyBE) -- (\lx, \lyBE);	
		\draw (\lxM,        \lyBE) -- (\lxM,        \yPCM);
		\draw (\lxM + \lxH, \lyBE) -- (\lxM + \lxH, \yPCM);
		\draw [dashed] (\lxNAc,         \yPCM) -- (\lxNAc,         \yTE);
		\draw [dashed] (\lxNAc + \lxAc, \yPCM) -- (\lxNAc + \lxAc, \yTE);

		\node at (\lx/2, \ly - \lyTE*.35) {Top electrode};
		\node at (\lx/2, \yPCM + \lyPCM/2) {GGST};
		\draw    (\lx/2, \lyBE + \lyMH*0.55) -- (\lx*.55, \lyBE + \lyMH*0.8)
			node [right] {Heater};
		\node at (       \lxM/2, \lyBE + \lyMH*.5) {Oxide};
		\node at (\xHD + \lxM/2, \lyBE + \lyMH*.4) {Oxide};
		\node at (\lx/2, \lyBE/2) {Bottom electrode};
		\draw [decorate, decoration = {brace, raise=2.5pt, amplitude=5pt}]
			(\lxNAc, \yTE) --  (\lxAc + \lxNAc, \yTE) node [pos=0.5, above=7pt] {Active GGST};

		\pgfdeclarelayer{bg}    
		\pgfsetlayers{bg,main}  
		\begin{pgfonlayer}{bg}
			\fill[yellow!40] (0,    \lyBE) rectangle (\lxM, \yPCM);
			\fill[red!50]    (\lxM, \lyBE) rectangle (\xHD, \yPCM);
			\fill[yellow!40] (\xHD, \lyBE) rectangle (\lx,  \yPCM);
			\fill[orange!50] (0, 0)     rectangle (\lx, \lyBE);	
			\fill[blue!30]   (0, \yPCM) rectangle (\lx, \yTE);		
			\fill[orange!50] (0, \yTE)  rectangle (\lx, \ly);		
		\end{pgfonlayer}

		\def\eDim{0.3};		
		\def\eFl{0.04};		
		\def\eN{2pt};		
		\draw [<->] (-\eDim, 0) -- (-\eDim, \ly)
			node [pos=0.5, left=\eN] {$\qty{240}{\nano\meter}$};
		\draw [<->] (0, \ly + \eDim) -- (\lx, \ly+ \eDim)
			node [pos=0.5, above=\eN] {$\qty{300}{\nano\meter}$};
		\draw[white,<->] (\lx + 0.9, \yTE + \eFl) -- (\lx + 0.9, \ly - \eFl);

		\end{tikzpicture}
	\caption{Simulation domain for memory operations.}
	\label{fig_schema_memoire}
\end{figure}
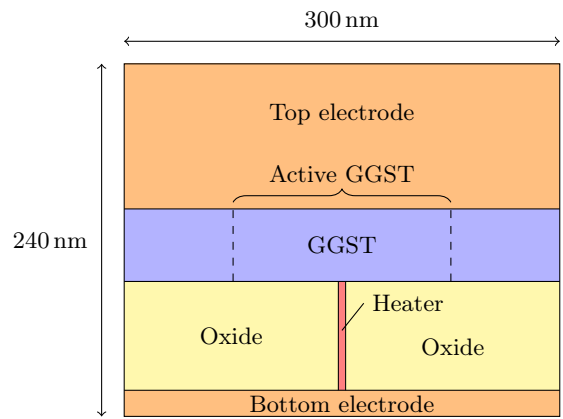

The geometry of PCM devices has been optimized to limit power consumption while operating 
the memory cell.
The so-called Wall architecture, schematically shown in Fig.~\ref{fig_structure_wall}, was found to maximize the current density inside the active volume, and thus to reduce the programming current.
While the exact geometric details of the structure are not important for our purpose, the main 
features should be reproduced to make a comparison with experimental measurement meaningful.
Therefore, we set up a two-dimensional simulation cell which represents a cross-section
through the wall structure, shown in Fig.~\ref{fig_schema_memoire}.

The electric currents that are used to probe or to change the state of the memory 
cell flow between a high and narrow electrode shown at the bottom of the figure, 
the so-called \emph{heater}, and an extended counterelectrode on the top. The Joule 
heating used to melt the material mainly occurs in the heater, but also inside the 
active material. The generated heat is evacuated to the surroundings of the memory 
cell by heat conduction.
A complete modeling of memory operations must therefore include equations for the
electric current, for heat generation and dissipation and for the structural changes
in the active material. Whereas the latter need to be solved only within the active
material, electric current and heat flow need also to be taken into account outside of 
the GGST layer. In order to optimize the performance of our code, we solve the different
equations on different domains with different simulation grids, as will be described
in detail below in Sec.~\ref{sec_numerics}.

The domain must be large enough so that the temperature field is not significantly disturbed by boundary conditions.
A total domain of \qty{300}{\nano\meter} width by \qty{240}{\nano\meter} height is used as a good trade-of between resolution speed and realistic thermal profile.

\subsection{Crystallization model} \label{ss_model_cryst}
All the details of the crystallization model have been presented in a previous 
publication \cite{bayle2020} and need not be repeated here; for reference, the 
model equations and parameters are summarized in appendix A.

The crystallization model has been formulated as a multi-phase field model (MPFM).
In this mesoscopic approach, the simulation domain can be occupied by different 
phases which are separated by diffuse interfaces. To capture the segregation effect, 
three phases are considered: crystalline germanium (Ge), crystalline \ce{Ge2Sb2Te5} (GST) 
and a disordered phase corresponding either to amorphous or to liquid GGST.
Each phase is associated with a scalar phase field $p_i$ ($i=1$ for Ge, $i=2$ for GST and $i=3$ for the disordered phase) that ranges between 0 and 1 (1 when the phase is present and 0 when it is not).

In addition, a concentration field $c$ is introduced, which corresponds to the local 
excess of germanium with respect to GST composition, meaning that $c=0$ is pure GST 
and $c=1$ is pure germanium.
This amounts to a pseudobinary approximation which restricts
the possible compositions of the ternary mixture to the straight segment linking
\ce{Ge2Sb2Te5} and pure Ge in the ternary phase diagram.

Interface motion and chemical diffusion are driven, respectively, by grand potential
differences between the phases and by gradients of the chemical potentials. Once the
free energy functions for all the phases are specified, the equations of motion can
be derived from the thermodynamics of irreversible processes.
Finally, an orientation field model is also included to account for the polycrystalline 
aspect of the GGST material and the formation of grain boundaries.
This adds two scalar orientation fields to the crystallization model, but 
they do not couple to the thermal and electrical models presented below.

\subsection{Thermal model}
The thermal model should reproduce the heat generation and its propagation in the different 
materials constituting the memory.
The temperature $T$ in the structure is computed using the Fourier equation:
\begin{equation}
	\frac{C_m}{V_m} \frac{\partial T}{\partial t} = \vv{\nabla} \cdot \left(k_{th}\vv{\nabla}T\right) + S
	\label{eq_fourier},
\end{equation}
with $k_{th}$ the thermal conductivity (in \unit{\kth}), $C_m$ the molar specific heat (in \unit{\cpMol}), $V_m$ the molar volume (in \unit{\meter\cubed\per\mole}) and $S$ a heat
source term (in \unit{\watt\per\meter\cubed}). To be consistent with the MPFM model,
we consider identical molar volumes for the three phases of the PCM material.
This approximation, commonly used in phase-field models for crystallization, implies 
that we do not take into account mechanical effects on the phase-change mechanism.

The multi-phase aspect inherent to the coupling between MPFM and thermal model is reflected 
in the expression for the thermal conductivity, which reads
\begin{equation}
	k_{th}(T, c) = g_1 \, k_{th}^{Ge}(T) + g_2 \, k_{th}^{GST}(T) + g_3 \, k_{th}^{A/L}(T,c),
	\label{eq_kth123}
\end{equation}
where the superscript $A/L$ stands for amorphous/liquid (the disordered phase), and
$g_i$ are weight coefficients associated with each phase, their sum being equal to 1 \cite{bayle2020}.
The additional dependence of $k_{th}^{A/L}$ on the concentration $c$ will be commented in Section \ref{sec_param} below.
Note that we have taken the specific heat to be constant and identical for the three phases.

The source term $S$ contains the contribution of latent heat generated by phase change, and the Joule heating produced by the electric current:

\begin{equation}
	S = \underbrace{
	    - \frac{L_{Ge}}{V_m} \frac{\partial g_1}{\partial t}
	    - \frac{L_{GST}}{V_m}\frac{\partial g_2}{\partial t}
	    }_{\text{Latent heats of phase change}}
	    +  \underbrace{\sigma\big( \vv{\nabla}V \big)^2 }_{\text{Joule heating}}
	\label{eq_source}
\end{equation}
with $L_{Ge}$ and $L_{GST}$ the latent heats of the two crystalline phases (in \unit{\joule\per\mole}), $V$ the electrostatic potential, and $\sigma$ the electrical conductivity, which will be detailed below.

In addition, as already mentioned in the introduction, considering the intensity of Joule heating observed in PCM devices with respect to their small size, thermal boundary resistances (TBR) are taken into account at the interfaces of the memory element interfaces (see Fig. \ref{fig_schema_memoire}, for instance at the two oxide/heater interfaces).
When TBR are introduced, the heat flux density through an interface, $\phi_i$, is related to a temperature difference $\Delta T$ between the two sides of the interface by $\phi_i = \Delta T/R_{i}$ with $R_{i}$ the thermal resistance of the interface. Under the conditions of typical memory
operations, this can lead to temperature drops of a few hundred kelvins.
The TBR are thus key to accurately reproduce the thermal confinement in the memory \cite{durai2020}.
TBR are integrated in the thermal conductivity term of Eq. \eqref{eq_fourier} in the following way.
In a discretized model, when an interface passes between two grid points distant by the grid 
spacing $dx$, of which one is in material A and the other in material B, the effective thermal conductivity $k_\text{eff}$ verifies
\begin{equation}
k_\text{eff} = \cfrac{1}{\cfrac{f}{k_{th}^A} + \cfrac{R_i}{dx} + \cfrac{1-f}{k_{th}^B}}
\label{eq_kth_TBR}
\end{equation}
with $f$ the normalized distance between the first node and the interface ($1-f$ for the second node), $R_{i}$ the thermal resistance of the interface, and $k_{th}^A$ and $k_{th}^B$ the thermal conductivities of the two materials.

\subsection{Electrical model} \label{ss_model_electrical}
Electrical conduction through the structure allows to perform both programming (via Joule heating) 
and reading operations (via resistivity sensing). The electrical potential is computed using 
the Laplace equation, which means that we assume the material to be charge neutral everywhere.
While this is fully justified for a conductor, germanium and GST are semiconductors \cite{kato2005},
and therefore in principle charge separation could locally occur.
We assume that the redistribution of free carriers is much faster than the other phenomena at play 
(heat conduction and crystallization), and that we can therefore consider 
a steady-state form of the current equation even in the PCM material.
This approximation, widely used in PCM device simulations \cite{baldo2020}, is adapted to simulate Joule heating and electric current paths in the microstructure generated by the MPFM.
\begin{equation}
\vv{\nabla} \cdot \vv{j} = 0 \quad \text{with} \quad \vv{j} = -\sigma \vv{\nabla} V,	\label{eq_laplace} \\
\end{equation}
with $\vv{j}$ the current density. In the PCM material, a phase-dependent electrical conductivty is used:
\begin{equation}
\sigma(T, E) = g_1 \, \sigma_{Ge}(T) + g_2 \, \sigma_{GST}(T) + g_3 \, \sigma_{\!A/L}(T, E).	\label{eq_sigma(T)}
\end{equation}

The additional dependence of $\sigma_{\!A/L}$ on the modulus of the electric field $E$ is important to reproduce the ovonic threshold switching effect \cite{pirovano2004} that is key to PCM operations:
at high electric field, the conductivity of the amorphous phase strongly increases to reach 
a value similar to the one of the crystalline material. This makes it possible to use 
programming pulses of a constant intensity, regardless of the previous memory state.

\section{Parameters} 
\label{sec_param}
In addition to the latent heat and molar volume that were already needed for the crystallization model \cite{bayle2020} (see Table \ref{tab_diag_param} and \ref{tab_autres_param}), the thermal and the electrical models require the knowledge of multiple physical parameters: thermal conductivity, specific heat, thermal boundary resistances, and electrical conductivity.
In the following, we review the data that can be gathered from the literature and describe our choices for the parametrization of our model.
For some parameters, measurement done internally on GST and GGST are also exploited.

\subsection{Thermal conductivities and specific heats in the three phases} \label{ss_param_kth_cp_GGST}
The thermal conductivities of the three phases (Ge, GST and amorphous/liquid) in Eq. \eqref{eq_kth123} range over two orders of magnitude.
For germanium, the temperature dependence from [\onlinecite{glassbrenner1964}] is approximated by several linear segments over the 300-\qty{1200}{\kelvin} range.
They can be constructed from the values listed in Table \ref{tab_kth_Ge}.
Those values serve as a starting point but they are reduced by a factor of 8 to match the lower thermal conductivity in germanium thin films \cite{wang2011}, see discussion in Sec.~\ref{ss_calib_kthGGST}.
\begin{table}[h!]
	\normalsize
	\caption{Values needed to construct $k_{th}^{Ge}(T)$.}
	\label{tab_kth_Ge}
	\centering
	\begin{tabular}{@{} lccccc @{}} \toprule
	$T$ (\unit{\kelvin}) 			& 299 	& 393	& 676	& 868	& 1171 	\\
	$\kth$ (\unit{\kth})			& 60.2	& 42.6	& 22.0	& 18.0	& 17.5 	\\ \bottomrule
	\end{tabular}
\end{table}

For GST, measurements from \num{425} to \SI{575}{\kelvin} \cite{lyeo2006} are considered and extrapolated linearly up to the temperature where the melting of the phase occurs:
\begin{equation}
	\kth^{GST} (T) = \num{2.94e-3} \times T - 0.806,
	\label{eq_kth_GST}
\end{equation}
with $T$ in kelvin.
They correspond to the face centered cubic (FCC) metastable structure of GST, the one observed experimentally \cite{agati2019}.
At lower temperature, a floor value of \qty{0.57}{\kth} is used, in agreement with [\onlinecite{lyeo2006}]. \vbl

Finally, for the common Ge-rich GST amorphous/liquid phase, two dependencies are considered:
one for the amorphous at low temperatures and one for the liquid at high temperatures.
They are connected linearly over a range of \qty{50}{\kelvin} centered at \qty{790}{\kelvin} (near the melting temperature, as in [\onlinecite{crespi2014}]).
A constant low thermal conductivity is considered for the amorphous \cite{kusiak2022}, $\kth^{A} = \qty{0.28}{\kth}$.
For liquid GGST, no experimental data are available.
As an alternative, liquid GST and liquid germanium values are combined as follows.
On the one hand, values for liquid GST coming from molecular dynamics 
simulations \cite{baratella2022} are approximated linearly by
\begin{equation}
	\kth^{GST,L} (T) = \num{2.4e-2} \times T - 18.8,
	\label{eq_kth_GST_L}
\end{equation}
with $T$ in kelvin.

\begin{figure}[!ht]
	\centering
	\includegraphics[width=0.44\textwidth]{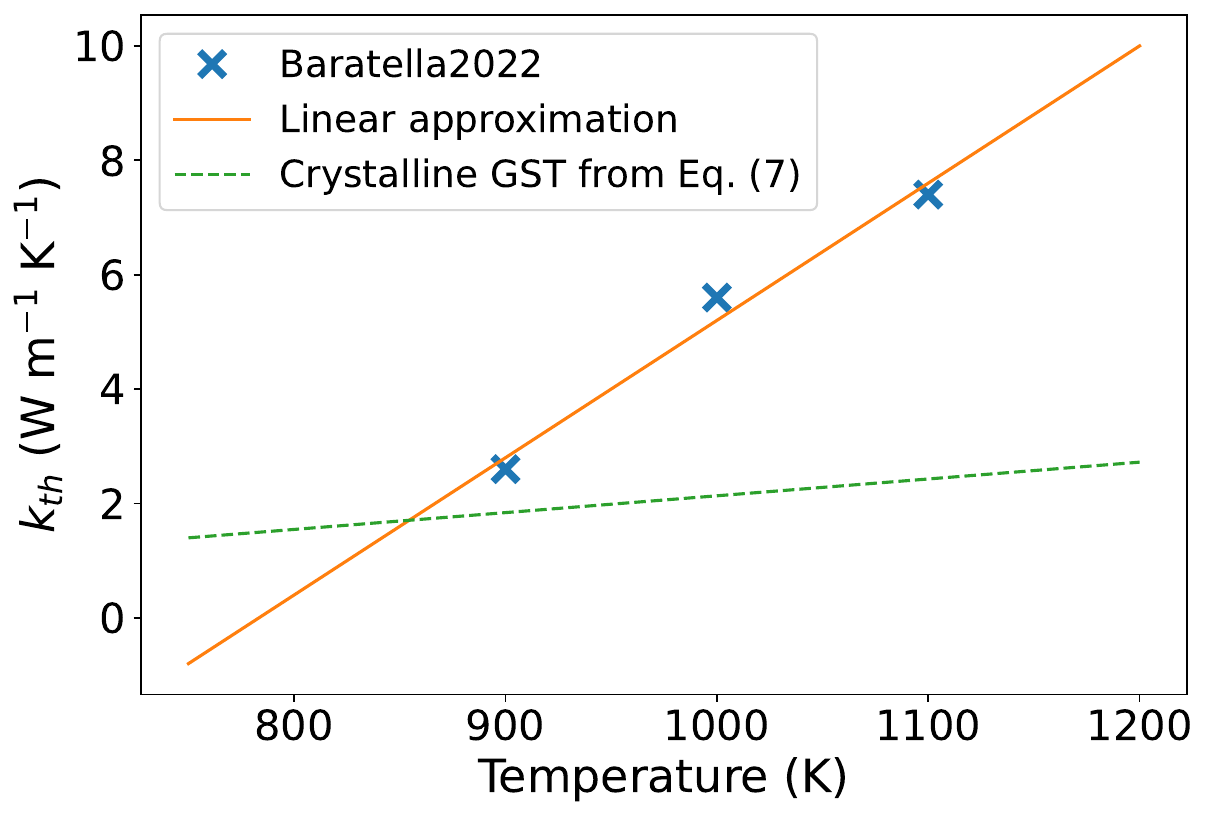}
	\caption{Thermal conductivity values from [\onlinecite{baratella2022}], a linear approximation, and the curve corresponding to Eq. \eqref{eq_kth_GST} for crystalline GST.
	This figure shows that the linear approximation of liquid GST values becomes negative below \qty{780}{\kelvin} and is lower than Eq. \eqref{eq_kth_GST} below \qty{850}{\kelvin}.}
	\label{fig_kth_GSTliq}
\end{figure}

To avoid unwanted negative values (that would appear around \qty{780}{\kelvin}), this temperature dependence is replaced by Eq. \eqref{eq_kth_GST} for temperatures where Eq. \eqref{eq_kth_GST} gives higher values than Eq. \eqref{eq_kth_GST_L}  (below \qty{850}{\kelvin}, as shown in Fig. \ref{fig_kth_GSTliq}).
\begin{figure*}[!ht]
	\centering
	\subfloat[Germanium]{
		\centering
		\includegraphics[height=5.5cm]{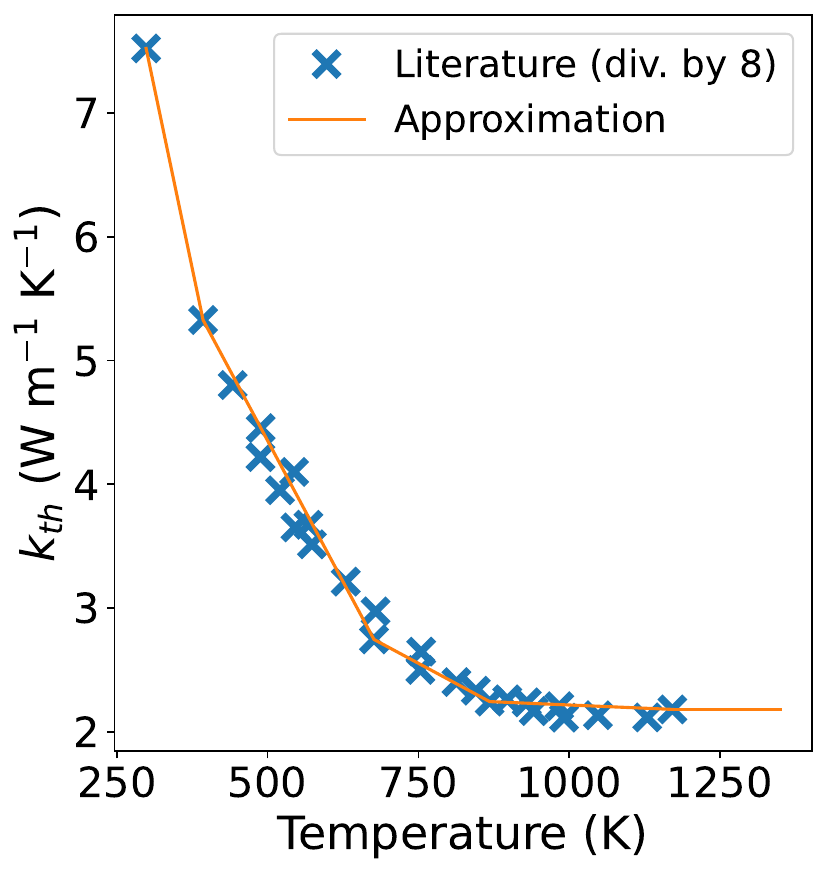}
	}
	\subfloat[GST]{
		\centering
		\includegraphics[height=5.5cm]{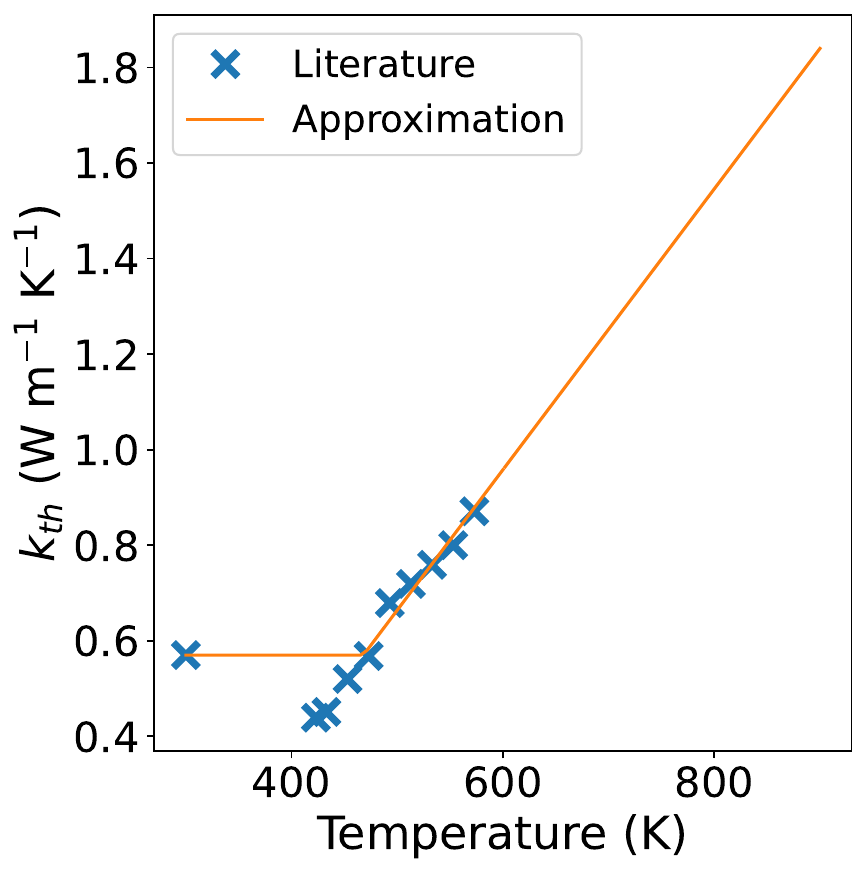}
	}
	\subfloat[Amorphous and liquid]{
		\centering
		\includegraphics[height=5.5cm]{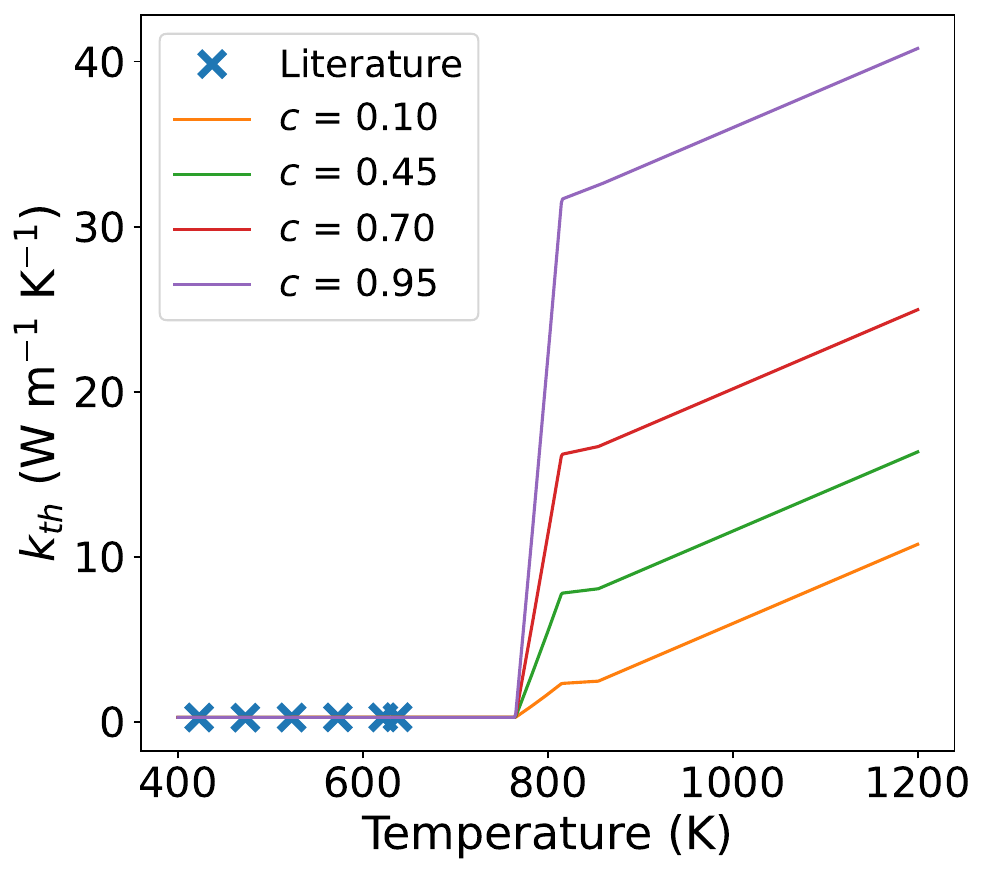}
	}
	\caption{Thermal conductivities versus temperature in the three phases, compared with relevant data from the literature.
	For germanium, literature data from [\onlinecite{glassbrenner1964}] are also divided by 8 to have a meaningful comparison.}
	\label{fig_kth_3_phases}
\end{figure*}
On the other hand, for liquid germanium, the following temperature dependence is used \cite{assael2017}:
\begin{equation}
	\kth^{Ge,L} (T) = 45.6 + 0.024(T - 1211),
\end{equation}
with $T$ in kelvin.
To combine them, the Filippow equation is used \cite{poling2001book}:
\begin{equation}
	\kth^{L} = w_xk_x + w_yk_y - 0.72w_xw_y(k_y-k_x),
\end{equation}
with $w_i$ the mass fractions, and $k_i$ the thermal conductivities; $x$ and $y$ corresponding respectively to GST and germanium.
The mass fractions are obtained using molar masses (\qty{72.6}{\gram\per\mole} for germanium and \qty{114.1}{\gram\per\mole} for GST) and they depend on the concentration of the liquid $c$.
This yields different curves for different values of $c$, which explains the dependence of $\kth^{A/L}$ on $c$.
Fig. \ref{fig_kth_3_phases} shows the thermal conductivites in the three phases, highlighting the large differences between each of them. \vbl

For the specific heat, the same constant value is used in the three phases, $C_m = \qty{26.7}{\cpMol}$.
It is close to the Dulong-Petit law and agrees with literature data available for germanium \cite{smith1966, chen1969} and GST \cite{kalb2002Thesis, scott2020} (less than \qty{20}{\percent} error).
This value should hold well for amorphous GGST, but it is probably underestimated for liquid GGST \cite{kalb2002Thesis}.
Nevertheless, as a first approach, we have chosen to also use it for the liquid.

\subsection{Thermal conductivities and specific heats in other materials}
The surroundings of the GGST layer are made of other materials. For simplicity, constant
values of $k_{th}$ and $C_m$ are used in these regions;
the values used are summarized in Table \ref{tab_kth_Cp_autres}.
\begin{table}[!b]
	\normalsize
	\centering
	\caption{Thermal conductivities (in \unit{\kth}) and specific heats (in \unit{\cpMol}) for additional materials.}
	\label{tab_kth_Cp_autres}
	\begin{tabular}{@{} llrr @{}} \toprule
						& Material 				& $\kth$		& $C_m$\\
	\midrule
	Top electrode		& \ce{TiN}				& 25.7		& 18.5 \\
	Bottom electrode		& \ce{W}					& 170 		& 24.2 \\
	Oxide				& \ce{Si3N4}				& 1.39 		& 17.9 \\
	Heater				& \ce{TiN} composite		& 13			& 23.6 \\
	\bottomrule
	\end{tabular}
\end{table}
For the top electrode, $\kth$ is provided in [\onlinecite{taylor1964}].
Due to the high Debye temperature of \ce{TiN}, the Dulong-Petit law does not hold and $C_m$ increases with temperature \cite{mohammadpour2018}.
However, during simulations, the material stays in the 300-\qty{400}{\kelvin} range, and therefore we use \qty{18.5}{\cpMol} (\qty{37}{\cpMol} read in [\onlinecite{mohammadpour2018}] on figure 4c, divided by two because there are two atoms per formula unit of \ce{TiN}).

The bottom electrode temperature is almost constant, too, because of the TBR and its high thermal conductivity; $\kth$ and $C_m$ come from [\onlinecite{white1997}].

Contrary to the two electrodes, the temperature in the oxide layers varies a lot.
$\kth$ of \ce{Si3N4} has been measured at CEA-Leti up to \qty{650}{\kelvin} and it increases linearly.
To keep the model simple, and since we do not have any information on the evolution at higher temperatures, we use the same constant value as other authors \cite{serra2019, kusiak2022}.
It corresponds to measurements made at \qty{560}{\kelvin}.
$C_m$ is also reported to increase with temperature \cite{NIST_JANAF1998book}.
For consistency, we use the value at \qty{560}{\kelvin}, \qty{125.4}{\cpMol}, and divide it by the number of atoms.

The heater is made of a material close to \ce{TiN} \cite{ranica2021} with a higher electrical resistivity.
No data are available for this material, however, an article reports a decrease of the thermal conductivity of \ce{TiAlSiN} as the amount of \ce{AlSi} increases \cite{samani2013}.
To take this into account, we divide the \ce{TiN} thermal conductivity by 2.
For the specific heat, we could not find any reference in the literature, so we rely on data for \ce{TiN}.
Since the temperature of the heater stays quite high for most of the simulation, we consider a higher temperature range than for \ce{TiN}, 600-\qty{2000}{\kelvin}.

\subsection{Thermal boundary resistances}
As for $k_{th}$ and $C_m$ in additional materials, TBR are taken constant but they vary depending on the materials on both sides of the interface.
The values used are summarized in Table \ref{tab_TBR}.
\begin{table}[!ht]
	\normalsize
	\centering
	\caption{Thermal boundary resistances (in \unit{\kelvin\meter\squared\per\giga\watt}).}
	\label{tab_TBR}
	\begin{tabular}{@{} llr @{}} \toprule
	Material 1			& Material 2				& $R_i$ \\
	\midrule
	Amorphous 		& \ce{TiN}/heater		& 210 \\
	Crystalline		& \ce{TiN}/heater		&  25 \\
	Liquid			& \ce{TiN}/heater		&  10 \\
	\midrule
	Amorphous		& \ce{Si3N4}				&  50 \\
	Crystalline		& \ce{Si3N4}				&   5 \\
	Liquid	 		& \ce{Si3N4}				&   2 \\
	\midrule
	\ce{Si3N4}		& \ce{TiN}				&   5 \\
	\ce{Si3N4}		& \ce{W}					&  15 \\
	\ce{TiN}			& \ce{W}					&   4 \\
	\bottomrule
	\end{tabular}
\end{table}
For the phase change material, we distinguish the amorphous, the crystalline and the liquid state.
The crystalline state correspond to both GST and germanium as both phases are not distinguished.
Indeed, most data on TBR at interfaces with \ce{TiN}, \ce{Si3N4} or the heater correspond to amorphous and crystalline GST (and not GGST).
Also, since no data were found for the interface with the heater material, values corresponding to \ce{TiN} are used instead. \vbl

For interfaces between \ce{TiN} and GGST, the value for the amorphous phase comes from [\onlinecite{lee2013}] and for the crystalline phase from [\onlinecite{lee2013}] and [\onlinecite{reifenberg2010}].
For \ce{Si3N4}, an author measures the same value for interfaces with amorphous and with crystalline GST \cite{kusiak2022}.
This is inconsistent with what is reported for \ce{TiN} and \ce{SiO2} \cite{battaglia2010}; the resistance at an interface with an amorphous phase should be higher.
As a consequence, for crystalline GGST we use the lower resistivity value from [\onlinecite{battaglia2010}].
TBR at interfaces with a liquid being impossible to measure, we use a value roughly 
divided by 20 with respect to the amorphous phase for both \ce{TiN} and \ce{Si3N4}.
The reasoning behind this choice is that we see an increase of the same order of 
magnitude in the thermal conductivity of GST when the material melts.

For \ce{Si3N4}/\ce{TiN} and \ce{Si3N4}/\ce{W} interfaces, an article finds a correlation between dielectric/metals TBR and the difference of the Debye temperatures of the two materials \cite{jeong2012}.
Debye temperature of tungsten, \ce{TiN} and \ce{Si3N4} are \qty{400}{\kelvin} \cite{kittel2004book}, \qty{900}{\kelvin} \cite{mohammadpour2018} and \qty{1150}{\kelvin} \cite{jiang2002}, respectively.
This yields the two values reported in Table \ref{tab_TBR}.
Finally, for the \ce{TiN}/\ce{W} interface, data corresponding to \ce{TiN}/\ce{Al} are used as reference \cite{reifenberg2010}. \vbl

The large uncertainties associated with the reported values (see [\onlinecite{lee2013}] for instance) and the general lack of data on those parameters make them more prone to significant errors.
Therefore, even though some articles report on temperature dependencies of the TBR, we chose to consider only constant values.

\subsection{Electrical conductivity} \label{ss_elconductivity}
Contrary to thermal parameters, electrical conductivities were mostly provided by unpublished in-house material characterization.
Since the electrical conductivities of Ge and GST are very
sensitive to the presence of impurities, for a meaningful comparison between simulations and experiments,
it is necessary to use the values measured on the material that is actually used in the devices.
In the three phases present in the active material, the main dependence considered for electrical conductivities is the temperature dependence. 
In addition, in the amorphous/liquid phase, several additional effects are included, as described below.

The electrical conductivity of the heater is taken as constant and has been calibrated to \qty{5e4}{\sig} (see discussion in Sec.~\ref{ss_calib_heater} below). \vbl

\begin{figure}[!ht]
	\centering
	\includegraphics[width=0.44\textwidth]{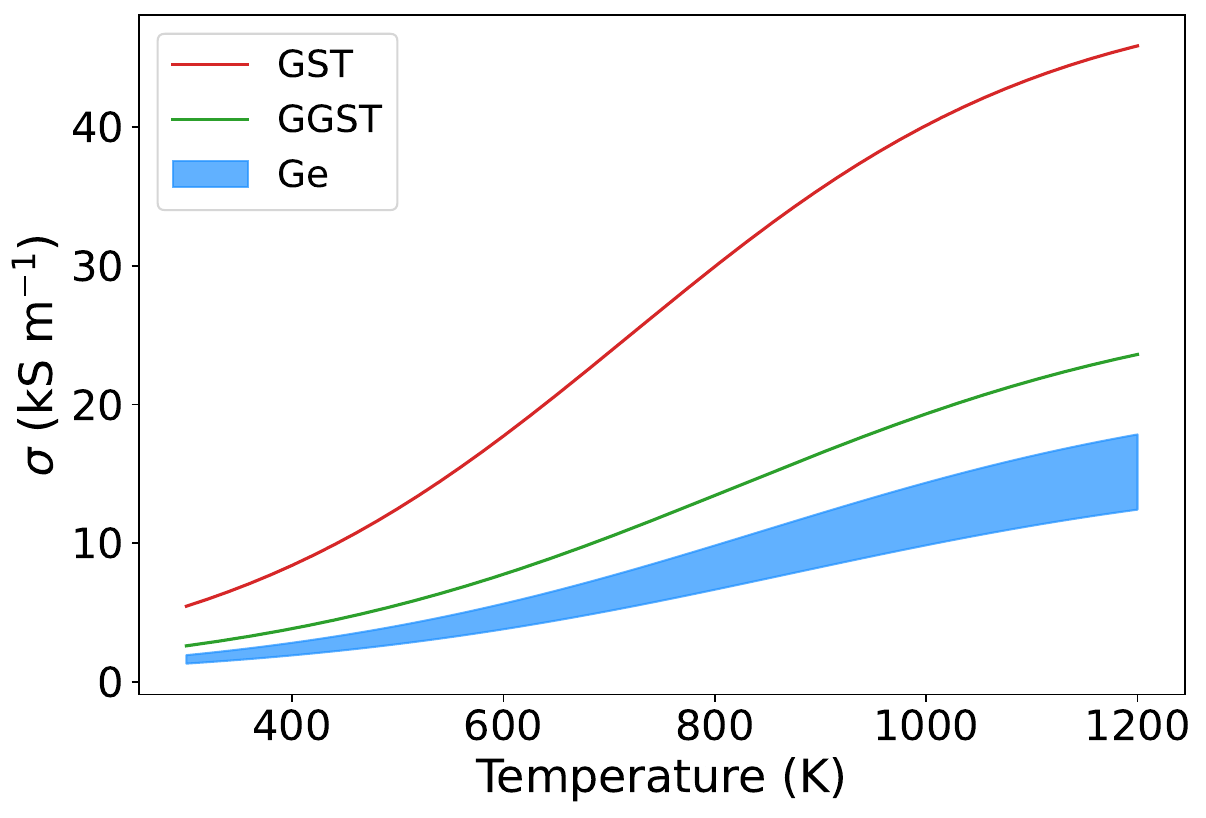}
	\caption{Electrical conductivities of GGST, GST and germanium.}
	\label{fig_sigmaGeGST}
\end{figure}

For germanium and GST, the two crystalline phases, we exploited measurements done at CEA-Leti on GST and GGST.
The resulting resistivities are fitted by hyperbolic tangents:
\begin{equation}
	\sigma (T) = \frac{a}{2} \big[ \tanh ( bT + c) + d \big],
	\label{eq_sigmaTanh}
\end{equation}
with coefficients listed in Table \ref{tab_sigma_coeff_tanh}.
\begin{table}[!ht]
	\normalsize
	\centering
	\caption{Coefficients of hyperbolic tangents for GST and GGST.}
	\label{tab_sigma_coeff_tanh}
	\begin{tabular}{@{} lcccc @{}} \toprule
					& $a \; [\unit{\sig}]$			& $b \; [\unit{\kelvin}^{-1}]$		& $c$	& $d$ 	\\
	\midrule
	GST 			& \num{5.0e4} 	& 0.0025	& -1.8	& 1.0	\\
	GGST 			& \num{2.8e4} 	& 0.0022	& -1.8	& 1.0	\\
	\bottomrule
	\end{tabular}
\end{table}
From those values, the electrical conductivity of germanium can be determined.
By following the same reasoning as for Eq. \eqref{eq_kth_TBR}, we can approximate GGST values by combining germanium and GST ones, and by accounting for the initial concentration of GGST $c_0$:
\begin{equation}
	\frac{ 1 }{ \sigma_{GGST}(T) } = \cfrac{ c_0 }{ \sigma_{Ge}(T) } + \cfrac{ 1 - c_0 }{ \sigma_{GST}(T) }.
	\label{eq_sigmaGe}
\end{equation}
$\sigma_{Ge}(T)$ is obtained by inverting this equation and the three curves are plotted in Fig. \ref{fig_sigmaGeGST}.
To avoid disclosing the exact value of $c_0$, which is confidential, a range is provided instead. The range is bounded by $c_0 = 0.35$ and $c_0 = 0.6$.

\vbl For the common amorphous/liquid phase, we first focus on the amorphous electrical conductivity. 
The ovonic threshold switching effect (see Sec.~\ref{ss_model_electrical}) is reproduced by using two temperature dependencies, one at low electric field and one at high electric field.
\begin{equation}
	\sigma_{\!A/L}(T, E) = \left\{
	\begin{array}{cl}
		\sigma_{\!A/L}^{low}(T) 	\quad &\text{ if }  E < E_{th} \\[0.2cm]
		\sigma_{\!A/L}^{high}(T)	\quad &\text{ if }  E \geq E_{th}
	\end{array} \right.
\end{equation}
The conductivity in the amorphous GGST for low electric field is modeled through the following equation:
\begin{equation}
	\sigma_{\!A/L}^{low}(T) = \sigma_0 \exp \left( - \frac{ E_A }{ k_B T } \right),
	\label{fig_sigmaAmGGST}
\end{equation}
with $\sigma_0 = \qty{6600}{\sig}$ and $E_A = \qty{0.2}{\electronvolt}$ an activation energy.
Those parameters are fitted from data of in-house measurements.
The switching from low conductivity to high conductivity happens in a cell of the computational mesh if the electric field locally reaches $E_{th} = \qty{4e7}{\volt\per\meter}$.
This value of $E_{th}$ has been determined empirically. It is high enough to avoid the switching of small amorphous domes during read operations (at \qty{0.1}{\volt}), but low enough to switch large domes during writing operations (at more than \qty{1}{\volt}).
It is coherent with the threshold switching voltage $V_{th} = \qty{1.3}{\volt}$ found in the literature \cite{ciocchini2014}.
The conductivity model used for high electric field is the same as the one for crystalline GGST (see Eq.~\eqref{eq_sigmaTanh} and Table~\ref{tab_sigma_coeff_tanh}).

On top of the threshold switching effect, we also include a Poole-Frenkel conduction in the amorphous GGST for low electric field.
A dependence on the electric field is added to $\sigma_{\!A/L}^{low}$ and Eq. \eqref{fig_sigmaAmGGST} becomes
\begin{equation}
	\sigma_{\!A/L}^{low}(T,E) = \sigma_0 \exp \left(
		- \cfrac{ E_A - q \sqrt{ \cfrac{ q E }{ 8 \pi \varepsilon_0 } } }{ k_B T } \right),
\end{equation}
with $q$ the elementary charge and $\varepsilon_0$ the vacuum permittivity.
The effect of the Poole-Frenkel conduction is to decrease the energy barrier for moderate electric field, which increases the electrical conductivity of the material. The necessity of its addition will
be further discussed in Sec.~\ref{ss_calib_RI} below.

\begin{figure}[!ht]
	\centering
	\includegraphics[width=0.44\textwidth]{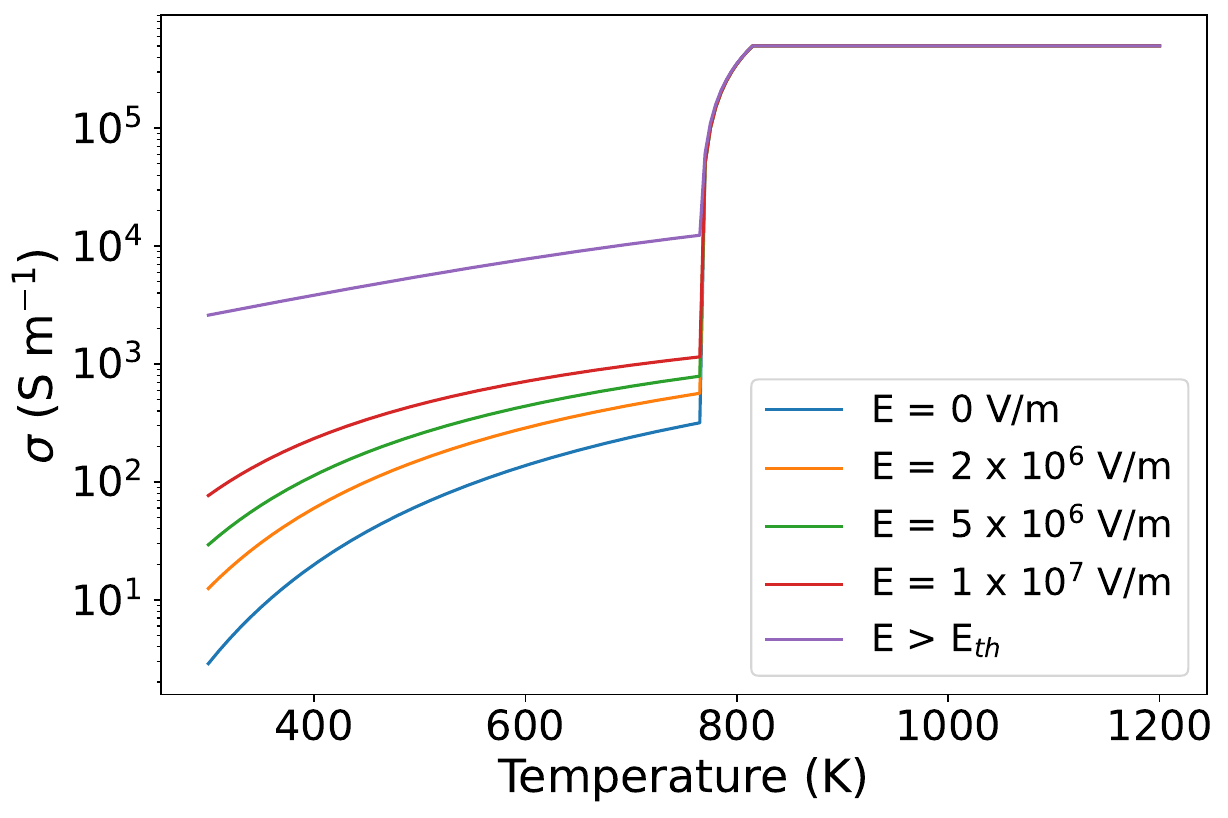}
	\caption{Electrical conductivities in the amorphous/liquid phase.
	Below \qty{790}{\kelvin}, the four curves at the bottom correspond to unswitched amorphous GGST and highlight the effect of the Poole-Frenkel conduction.
	The top curve, used above $E_{th}$ corresponds to crystalline GGST (see Eq. \eqref{eq_sigmaTanh} and Table \ref{tab_sigma_coeff_tanh}).
	At the melting temperature, all the curves increase to reach \qty{5e5}{\sig}.}
	\label{fig_sigmaAmLiq}
\end{figure}

Finally, above the melting temperature, the conductivity of the liquid phase is equal to \qty{5e5}{\sig}, as reported in [\onlinecite{crespi2014}].
As for the thermal conductivity, amorphous and liquid values are connected linearly.
Fig. \ref{fig_sigmaAmLiq} summarizes this information.

\section{Implementation} \label{sec_numerics}
\subsection{Simulation domain} \label{ss_implementation_domain}
While the thermal model is solved on the full domain, the two other models are solved on specific sub-parts.
The MPFM is solved in the ``active GGST'' region (see Fig. \ref{fig_schema_memoire}), the only area where the temperature rise is sufficient to trigger a phase change.
Its width has been chosen according to experimental observations and early simulation results.
Similarly, the electrical model is solved on a reduced domain comprised of this active region and of the heater.
Oxides are approximated as perfect insulators and the two electrodes as perfect conductors.

\begin{figure}[!ht]

	\newcommand{\drawDomain}{
		\def\scale{0.0107}		
		\def\lx{   250 * \scale}	
		\def\lxH{   10 * \scale}	
		\def\lxAc{ 120 * \scale}	
		\def\lyTE{ 100 * \scale}	
		\def\lyPCM{ 50 * \scale}	
		\def\lyMH{  75 * \scale}	
		\def\lyBE{  18 * \scale}	
		\pgfmathsetmacro\ly{\lyTE + \lyPCM + \lyMH + \lyBE}	
		\pgfmathsetmacro\lxM{(\lx - \lxH) / 2}		
		\pgfmathsetmacro\lxNAc{(\lx - \lxAc) / 2}	
		\pgfmathsetmacro\xHD{\lxM + \lxH}			
		\pgfmathsetmacro\yTE{\ly - \lyTE}			
		\pgfmathsetmacro\yPCM{\lyMH + \lyBE}		

		\draw (0, 0) rectangle (\lx, \ly);
		\draw (0, \yTE)  -- (\lx, \yTE);	
		\draw (0, \yPCM) -- (\lx, \yPCM);	
		\draw (0, \lyBE) -- (\lx, \lyBE);	
		\draw (\lxM,        \lyBE) -- (\lxM,        \yPCM);
		\draw (\lxM + \lxH, \lyBE) -- (\lxM + \lxH, \yPCM);
		\draw [dashed] (\lxNAc,         \yPCM) -- (\lxNAc,         \yTE);
		\draw [dashed] (\lxNAc + \lxAc, \yPCM) -- (\lxNAc + \lxAc, \yTE);
	}

	\centering
	\begin{tikzpicture}
		\def\dx{0.245}	

		\colorlet{cEltd}{orange!50}		
		\colorlet{cGGST}{blue!30}		
		\colorlet{cOxyd}{yellow!40}		
		\colorlet{cHeat}{red!50}		
		\def\op{0.0}	
		\definecolor{CL}{rgb}{.0, .65, .0}	

		\drawDomain

		\node [below=-0.55cm] at (\lx/2, \ly) {Crystallization};		

		\pgfdeclarelayer{bg}    
		\pgfsetlayers{bg,main}  
		\begin{pgfonlayer}{bg}
			\fill[cOxyd, opacity=\op] (0,    \lyBE) rectangle (\lxM, \yPCM);
			\fill[cHeat, opacity=\op] (\lxM, \lyBE) rectangle (\xHD, \yPCM);
			\fill[cOxyd, opacity=\op] (\xHD, \lyBE) rectangle (\lx,  \yPCM);
			\fill[cEltd, opacity=\op] (0, 0)     rectangle (\lx,  \lyBE);		
			\fill[cEltd, opacity=\op] (0, \yTE)  rectangle (\lx,  \ly);		
			\fill[cGGST, opacity=\op] (0,              \yPCM) rectangle (\lxNAc,         \yTE); 
			\fill[cGGST]              (\lxNAc,         \yPCM) rectangle (\lxNAc + \lxAc, \yTE); 
			\fill[cGGST, opacity=\op] (\lxNAc + \lxAc, \yPCM) rectangle (\lx,            \yTE); 
		\end{pgfonlayer}

		\draw [CL, line width=0.1cm](\lxNAc,         \yPCM) -- (\lxNAc,         \yTE);
		\draw [CL, line width=0.1cm](\lxNAc + \lxAc, \yPCM) -- (\lxNAc + \lxAc, \yTE);
		\draw [dashed]              (\lxNAc,         \yPCM) -- (\lxNAc,         \yTE);
		\draw [dashed]              (\lxNAc + \lxAc, \yPCM) -- (\lxNAc + \lxAc, \yTE);

		\tikzset{shift={(\lx + \dx, 0)}}

		\drawDomain

		\node [below=-0.55cm] at (\lx/2, \ly) {Thermal};

		\pgfdeclarelayer{bg}    
		\pgfsetlayers{bg,main}  
		\begin{pgfonlayer}{bg}
			\fill[cOxyd] (0,    \lyBE) rectangle (\lxM, \yPCM);
			\fill[cHeat] (\lxM, \lyBE) rectangle (\xHD, \yPCM);
			\fill[cOxyd] (\xHD, \lyBE) rectangle (\lx,  \yPCM);
			\fill[cEltd] (0, 0)     rectangle (\lx, \lyBE);	
			\fill[cGGST] (0, \yPCM) rectangle (\lx, \yTE);		
			\fill[cEltd] (0, \yTE)  rectangle (\lx, \ly);		
		\end{pgfonlayer}

		\draw [CL, line width=0.1cm](0, 0) rectangle (\lx, \ly);

		\tikzset{shift={(\lx + \dx, 0)}}

		\drawDomain

		\node [below=-0.55cm] at (\lx/2, \ly) {Electrical};

		\pgfdeclarelayer{bg}    
		\pgfsetlayers{bg,main}  
		\begin{pgfonlayer}{bg}
			\fill[cOxyd, opacity=\op] (0,    \lyBE) rectangle (\lxM, \yPCM);
			\fill[cHeat]              (\lxM, \lyBE) rectangle (\xHD, \yPCM);
			\fill[cOxyd, opacity=\op] (\xHD, \lyBE) rectangle (\lx,  \yPCM);
			\fill[cEltd, opacity=\op] (0, 0)     rectangle (\lx,  \lyBE);		
			\fill[cEltd, opacity=\op] (0, \yTE)  rectangle (\lx,  \ly);		
			\fill[cGGST, opacity=\op] (0,              \yPCM) rectangle (\lxNAc,         \yTE); 
			\fill[cGGST]              (\lxNAc,         \yPCM) rectangle (\lxNAc + \lxAc, \yTE); 
			\fill[cGGST, opacity=\op] (\lxNAc + \lxAc, \yPCM) rectangle (\lx,            \yTE); 
		\end{pgfonlayer}

		\def\d{0.05}	
		\draw [CL, line width=0.1cm] (\lxNAc,    \yTE)  -- (\lxNAc + \lxAc,   \yTE);	
		\draw [CL, line width=0.1cm] (\lxM - \d, \lyBE) -- (\lxM + \lxH + \d, \lyBE);	
	\end{tikzpicture}
	\caption{Resolution domains and boundary conditions associated with each model.
    Resolution domains are represented using plain colors and Dirichlet boundary conditions are in green.}
	\label{fig_domain_model}
\end{figure}
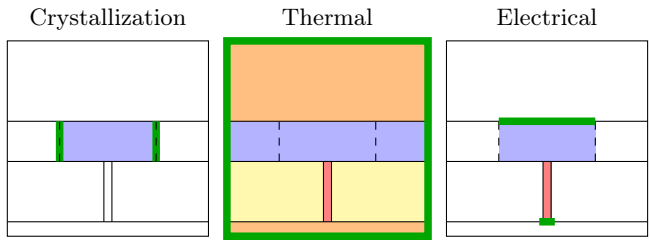

Boundary conditions also vary from one equation to another.
For the crystallization model, the continuity of the microstructure between active and non-active parts of the GGST region is maintained at the left and right interfaces.
For the temperature, Dirichlet boundary conditions are applied all around the domain and fixed at \SI{323}{\kelvin}.
Finally, for the electrical model, two potentials are applied on the top and the bottom electrodes (if a fixed current is desired, the potentials can be varied with time during the simulation), and zero-flux conditions are imposed at oxide interfaces and other domain boundaries.
A summary of the simulation domains and boundary conditions associated with each model is presented in Fig. \ref{fig_domain_model}.

\subsection{Numerical methods}
The three models are implemented in a simulator written in C++, using finite differences and an explicit (forward Euler) timestepping scheme.
In this framework, the maximal value of the simulation timestep $dt$ is set by the
fastest diffusive phenomenon considered according to the stability criterion
\begin{equation}
	dt \leq \frac{dx^2}{4D},
	\label{eq_critere_dt}
\end{equation}
with $D$ the diffusion coefficient in \unit{\meter\squared\per\second} ($D = \kth/C_m$ for the heat equation) and $dx$ the grid spacing.

The complete coupled model is a numerically stiff multi-scale problem. The characteristic
size of the grains and domains in the active material is a few nanometers. Therefore, the thickness 
of the diffuse interfaces of the phase-field model must be substantially smaller, that is, a
fraction of a nanometer. This, in turn, sets the maximal grid spacing that can be used, since
the interfaces must be resolved by several grid points. With this constraint, the thermal
diffusion inside the bottom electrode (made of tungsten) limits the timestep to $dt = \qty{e-16}{\second}$ 
(using the parameters in Table \ref{tab_kth_Cp_autres}). For the simulation of a memory
operation, which takes place on a microsecond time scale, the required number of timesteps
would be prohibitively large.

To overcome this limitation, we first tried to switch to an implicit (backward Euler) time scheme.
However, improvement made on the timestep were offset by the numerical cost of inverting a large matrix.
A much better efficiency gain was obtained by two improvements that introduce different grid spacings
and timesteps for the different model equations. \vbl

\begin{figure}[!ht]
	\centering
	\begin{tikzpicture}
		[perio/.style={gray, opacity=.2, rounded corners, fill}]

		\def\dxF{0.30};
		\def\rG{5};
		\pgfmathsetmacro\dxG{\rG * \dxF};
		\def\rPtF{0.02};
		\pgfmathsetmacro\rPtG{\rPtF * 6};

		\def\nxMG{5};
		\def\nyMG{3};

		\pgfmathsetmacro\xMax{\rG*\nxMG - 1};
		\pgfmathsetmacro\yMax{\rG*\nyMG - 1};
		\foreach \x in {0,...,\xMax}
		\foreach \y in {0,...,\yMax} \draw[fill] (\x*\dxF,\y*\dxF) circle [radius=\rPtF];

		\tikzset{shift={(2*\dxF, 2*\dxF)}}
		\pgfmathsetmacro\xMax{\nxMG - 1};
		\pgfmathsetmacro\yMax{\nyMG - 1};
		\foreach \x in {0,...,\xMax} {
			\foreach \y in {0,...,\yMax} \draw[red]    (\x*\dxG, \y*\dxG) circle [radius=\rPtG];
		}

	\end{tikzpicture}
	\caption{Refined grid (in black) and coarse grid (in red), with $r_G = 5$.
	Only a subset of the full domain is shown for better visibility.}
	\label{fig_multi_grille}
\end{figure}
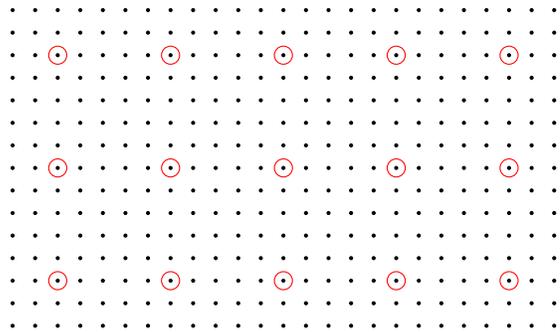

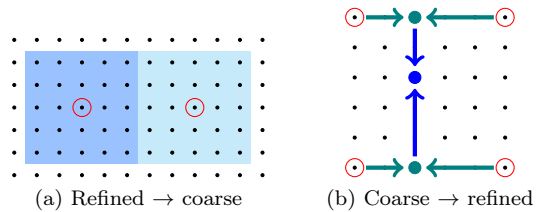
\begin{figure}[!ht]
	\centering

	\def\rG{5}
	\def\rPtF{0.02}
	\pgfmathsetmacro\rPtG{\rPtF * 6}

	\subfloat[Refined $\ra$ coarse]{
		\centering
		\begin{tikzpicture}
			\def\dxF{0.30};
			\pgfmathsetmacro\dxG{\rG * \dxF};
			\def\nxMG{2}
			\def\nyMG{1}

			\pgfmathsetmacro\xMax{\rG*\nxMG + 2 - 1};
			\pgfmathsetmacro\yMax{\rG*\nyMG + 2 - 1};
			\foreach \x in {0,...,\xMax}
			\foreach \y in {0,...,\yMax} \draw[fill] (\x*\dxF,\y*\dxF) circle [radius=\rPtF];

			\tikzset{shift={(3*\dxF, 3*\dxF)}}
			\pgfmathsetmacro\xMax{\nxMG - 1};
			\foreach \x in {0,...,\xMax} {
				\draw[red] (\x*\dxG, 0) circle [radius=\rPtG];
			}

			\pgfdeclarelayer{bg}    
			\pgfsetlayers{bg,main}  
			\begin{pgfonlayer}{bg}
				\fill[bluyan!40] (-\dxG*0.5, -\dxG*0.5) rectangle (\dxG*0.5, \dxG*0.5);
				\fill[cyan!20]   ( \dxG*0.5, -\dxG*0.5) rectangle (\dxG*1.5, \dxG*0.5);
			\end{pgfonlayer}
		\end{tikzpicture}
	}
	\quad
	\subfloat[Coarse $\ra$ refined]{
		\centering
		\begin{tikzpicture}
			\def\dxF{0.4};
			\def\dxG{\rG * \dxF}

			\foreach \x in {0,...,\rG}
			\foreach \y in {0,...,\rG} \draw[fill] (\x*\dxF, \y*\dxF) circle [radius=\rPtF];

			\foreach \x in {0,1}
			\foreach \y in {0,1} \draw[red]  (\x*\dxG, \y*\dxG) circle [radius=\rPtG];

			\draw[blue, fill] (2*\dxF, 3*\dxF) circle [radius=\rPtF*4];
			\draw[teal, fill] (2*\dxF, \dxG)   circle [radius=\rPtF*4];
			\draw[teal, fill] (2*\dxF, 0)      circle [radius=\rPtF*4];

			\def\esp{0.15};	
			\foreach \y in {0,\dxG} {
				\draw[->, teal, ultra thick] (\esp,        \y) -- (2*\dxF - \esp, \y);
				\draw[->, teal, ultra thick] (\dxG - \esp, \y) -- (2*\dxF + \esp, \y);
			}
			\draw[->, blue, ultra thick] (2*\dxF, \dxG-\esp) -- (2*\dxF, 3 * \dxF + \esp);
			\draw[->, blue, ultra thick] (2*\dxF, \esp)      -- (2*\dxF, 3 * \dxF - \esp);

			\draw[white] (-0.65, 0) -- (-0.6, 0);
			\draw[white] ( 2.65,  0) -- (2.6, 0);
		\end{tikzpicture}
	}
	\caption{Schematics summarizing how to pass fields from one grid to the other.}
	\label{fig_passage_grille}
\end{figure}

The first improvement is to solve the thermal model on a coarser grid with $dx' = r_G dx$, since $dx$ is only limited for the MPFM.
The gain on the timestep is significant ($dt \rightarrow dt \times r_G^2$) even for reasonably small values of $r_G$.
The position of the two grids, one relative to the other, can be seen on Fig. \ref{fig_multi_grille}.
The use of this method requires the transfer of data from one grid to the other.
For instance, the temperature is calculated on the coarse grid, but it is also needed
on the fine grid to evaluate physical parameters in the MPFM, such as the free energies.
Passing a field from the fine to the coarse grid is done by averaging the $r_G^2$ fine 
nodes that are closest to each coarse node. The transfer from the coarse to the fine grid 
is done with a double linear interpolation (to have every fine node surrounded by 
four coarse nodes, even at domain borders, we duplicate coarse nodes in accordance with the respective boundary conditions).
The two operations are shown in Fig. \ref{fig_passage_grille}.

One last important aspect of the two-grid implementation is related to the calculation of thermal conductivities on the coarse grid.
Indeed, thermal conductivities depend on the phases that are
locally present and must therefore be computed on the fine grid with Eq.~\eqref{eq_kth123}, 
while they are used on the coarse grid to calculate the thermal fluxes between nodes that
are needed to solve Eq.~\eqref{eq_fourier}. Applying the averaging procedure outlined above
directly to the thermal conductivity leads to large errors.
Instead, the underlying physics of the thermal conduction should be considered.
The heat flowing from one coarse node to the next actually passes through $r_G$ links (between $r_G+1$ nodes) on the fine grid.
The effective thermal conductivity $k_\text{eff}$ between the two coarse nodes therefore corresponds to those $r_G$ conductivities ``in series'':
\begin{equation}
\frac{1}{k_\text{eff}} = \dfrac{1}{N}\sum\limits_{i=0}^{N-1} \dfrac{1}{k_{i,i+1}},
\label{eq_keff_MG}
\end{equation}
with $k_{i,i+1} = (k_{i} + k_{i+1}) / 2$ the thermal conductivity between nodes $i$ and $i+1$ on the fine grid (Eq. \eqref{eq_kth_TBR} replaces $k_{i,i+1}$ at boundaries).

With this method, the actual improvement in computational efficiency is close to $r_G^2$, the expected gain from increasing the timestep.
During the evaluation of the method with $r_G$ values from 5 to 11, the simulation times were improved by \qty{88}{\percent} to \qty{100}{\percent} of $r_G^2$.
Therefore, the additional operations required to manage the two grids (see above) do not affect performances significantly.
\vbl

The second improvement is to solve the thermal model with a reduced timestep $dt_{th}$ compared to the crystallization model ($dt_{cr} = N_{dt} dt_{th}$).
The timestep of the thermal model has to satisfy the most stringent stability criterion, while the timestep of the crystallization model can be significantly higher.
This method implies to solve the thermal model multiple times between each resolution of the crystallization model.
Since the time scale of thermal conduction is much lower than the one of crystallization, it is a good approximation to consider that the fields of the MPFM ($p_i$, $c$...) do not evolve between each step of the thermal model.
Good performance improvements are obtained when it is much faster to solve the thermal model than the crystallization model (which is the case in our equation system). Indeed, compared to the reference 
case where both models are solved in each timestep, the former is solved the same number of times 
while the latter is solved $N_{dt}$ times less often.

Several $N_{dt}$ values were evaluated and the results are shown in Table \ref{tab_PS_time}.
For high $N_{dt}$ values, the speed-up reaches a maximum.
The is because the method is intrinsically limited by the ratio of simulation times for the thermal model and the crystallization model.
\begin{table}[!h]
	\normalsize
	\centering
	\caption{Simulation time speed-up for several $N_{dt}$ values.}
	\label{tab_PS_time}
	\begin{tabular}{@{} lcccccccccccc @{}} \toprule
	$N_{dt}$ 		& & 5	& & 10	& & 50	& & 100	& & 150 & &	750 \\	
	Speed-up\quad 	& & 4.1 & & 6.8	& & 15.1	& & 17.4	& & 19.0	& & 20.4 \\
	\bottomrule
	\end{tabular}
\end{table}

\begin{figure}[!ht]
	\centering
	\includegraphics[width=0.48\textwidth]{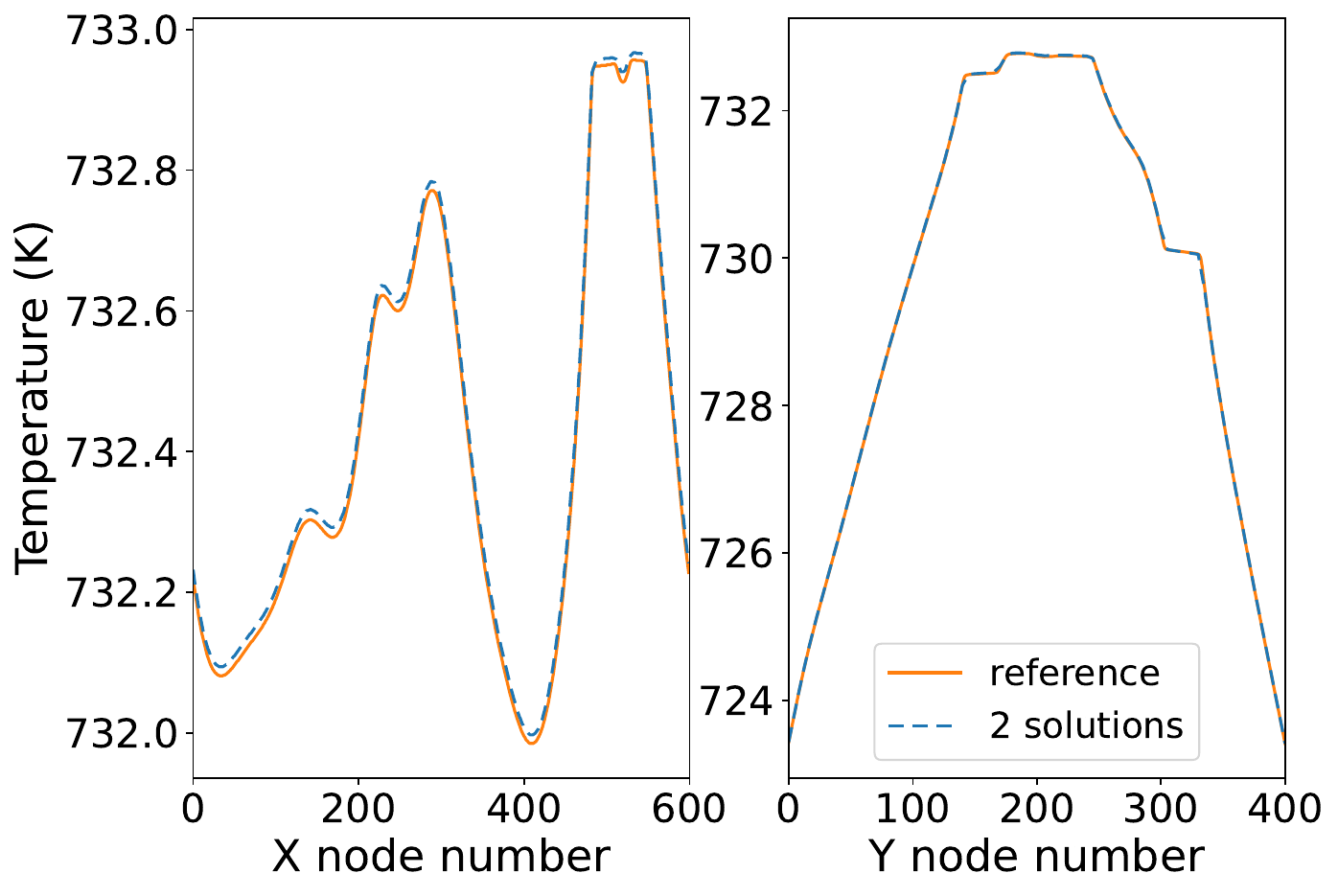}
	\caption{Comparison between temperature profiles of a reference solution and the result of the method with multiple grids and timesteps.
	The profiles come from annealing simulations in a 600 by 400 nodes domain.
	Left and right plots correspond to cuts along two directions (X and Y), at the center of the simulation domain.
	The differences between the two cases are less than 1\% of the total temperature increase.}
	\label{fig_temp_MG_PS}
\end{figure}

In addition to the improvements descrined above, we implemented parallelization with OpenMP (shared memory parallelization).
We tuned $r_G$ and $N_{dt}$ ($r_G = 5$ and $N_{dt} = 30$) to ensure an acceptable error on the results.
Fig. \ref{fig_temp_MG_PS} compares temperature profiles with a reference case where all the fields are
solved on the fine grid with the same timestep.
The differences between the two are negligible.
After all of these optimizations, most simulations run in only a few hours of wallclock time.

\subsection{Electrical model resolution and optimization}
Equation \eqref{eq_laplace} being stationary, the electrical model is not plagued by stability issues. However, it requires the solution of a matrix system, which can be computationally expensive.
To reduce the size of the matrix and to improve performance, the coarse grid presented above for the thermal model is also used for the electrical model.
In addition, early simulation results showed that the current density and the potential gradient are constant along the heater.
Consequently, the domain corresponding to the heater is reduced to one single horizontal line, instead of its entire length.
This approximation was tested against simulation on the fully discretized domain, and the error made on the electrical potential is negligible ($<\qty{1}{\percent}$).

To solve the matrix, the conjugate gradient method provided by the Eigen C++ library \cite{eigen} is used.
At each step, electrical conductivities are determined with the relevant fields (temperature and electric field) calculated in the previous timestep, and then the matrix is updated and the system is solved.
In section \ref{ss_implementation_domain}, it was already mentioned that the boundary conditions could be adjusted to force a fixed current instead of a fixed voltage.
To achieve this, after the first solution, the current is computed, and if it does not match the target value, one of the two voltage boundary conditions is changed and the system is solved again.

\subsection{Initial conditions} \label{ss_initial_cond}
For memory operation simulations, the initial conditions for the thermal and electrical models are simple.
The temperature is constant in the domain at \qty{323}{\kelvin}, and no electric current flows through the memory (the potentials at the two electrodes are the same).

\begin{figure}[!ht]
	\centering
	\includegraphics[width=0.34\textwidth]{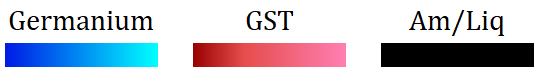} \vspace{0.12cm}\\
	\includegraphics[width=0.37\textwidth]{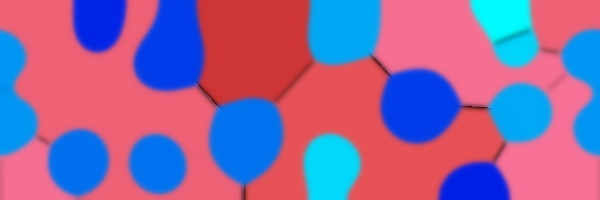}
	\caption{Initial microstructure used for most memory operation simulations.
	Germanium domains are blue, GST domains are red, amorphous and liquid domains are black.
	Different shades of blue and red mean different grain orientations.
	This figure corresponds only to the ``active GGST'' area (see Fig. \ref{fig_schema_memoire}).
	Exploiting its periodicity, it is duplicated to fill the two non-active part on each side.}
	\label{fig_initial_microstructure}
\end{figure}

The initial microstructure, on the other hand, is a fully crystallized layer of GGST (see Fig. \ref{fig_initial_microstructure}), with crystalline germanium and GST.
This microstructure is generated by an annealing simulation that aims at reproducing the crystallization of the material (and subsequent segregation) occurring during the fabrication process of the device.
As described in more detail in our previous work \cite{bayle2020}, this crystallized layer is obtained by annealing an amorphous domain (here, at \qty{723}{\kelvin}) in which germanium and GST grains nucleate and then grow using the crystallization model.
A noteworthy aspect is that during the nucleation process, crystalline seeds are placed randomly, such that different runs lead to different grain distributions.
The microstructure presented in Fig. \ref{fig_initial_microstructure} is used for most of the simulations presented in the next section.

\section{Results and discussion}
\label{sec_results}
\subsection{Outline}
The model contains a large number of parameters. Some of those (as for example the TBR) are
not known with great precision, others (as for example the electrical conductivities) can vary
with the presence of impurities. Therefore, not surprisingly, simulations with all the 
values of the parameters as gathered from the literature did not match the experimental data. 
However, guided by the comparison between simulations and experiments, we were able to 
refine some of our choices and thus to calibrate our model. In order to properly discuss
these adaptations, it is useful to have the main simulation results in mind. Therefore,
we will first present our main simulation results, obtained with the optimized parameter
set, and then discuss below (in Sec. \ref{ss_calibration}) all the changes that
have been made to obtain these results, and the reasoning behind these modifications.

\subsection{RESET operation}
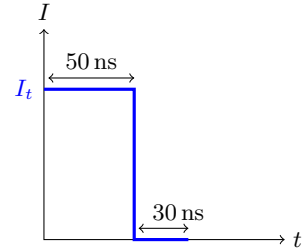
\begin{figure}[!ht]
	\centering
	\begin{tikzpicture}
		\def\tUp{1.2}							
		\pgfmathsetmacro\tDown{\tUp * 3/5}		
		\def\xMax{3.2}		
		\def\yMax{2.8}		
		\def\vPulse{2.0}	

		\draw [<->] (0, \yMax) node [above] {$I$} -- (0, 0) -- (\xMax, 0) node [right] {$t$};

		\draw [blue, very thick] (0,    \vPulse) node [left] {$I_t$}
		                       -- (\tUp, \vPulse) -- (\tUp, 0) -- (\tUp + \tDown, 0);

		\def\hFl{0.15}	
		\def\eFl{0.06}	
		\draw [<->] (\eFl,        \hFl + \vPulse) -- (\tUp, \hFl + \vPulse)
			node [pos=.5, above] {\qty{50}{\nano\second}};
		\draw [<->] (\eFl + \tUp, \hFl) -- (\tUp + \tDown,  \hFl)
			node [pos=.8, above] {\qty{30}{\nano\second}};
	\end{tikzpicture}
	\caption{Electrical pulse used for RESET simulations.}
	\label{fig_pulse_RESET}
\end{figure}

The programming pulse used during a RESET simulation (similar to the industry standard) is presented in Fig. \ref{fig_pulse_RESET}.
After the beginning of the pulse, the material heats up and, if the current density is high enough, it melts.
After \qty{50}{\nano\second}, the current is set to zero, and the memory cools down within \qty{30}{\nano\second}.

\begin{figure}[!ht]
	\centering
	\includegraphics[width=0.34\textwidth]{legende_microstructure.png} \vspace{0.12cm}\\
	\includegraphics[width=0.47\textwidth]{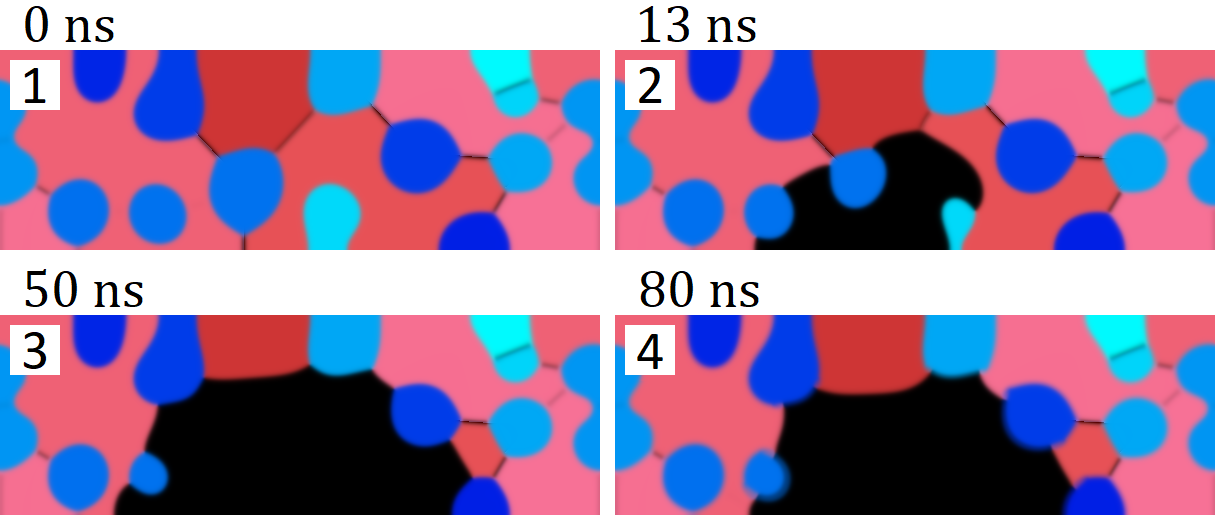}
	\caption{Microstructure evolution during a RESET pulse: before melting (1), during the pulse (2, 3), after operation (4).}
	\label{fig_reset_structure}
\end{figure}

The evolution of the microstructure during this simulation is displayed in Fig. \ref{fig_reset_structure}.
From the initial fully crystallized layer, a dome-shaped domain (in black) is first melted, then quenched by the fast decrease of the temperature.
Between steps 3 (at $t = \qty{50}{\nano\second}$) and 4 (end of simulation), a slight recrystallization begins but quickly stops as the temperature drops.
This is visible through the small reduction of the amorphous domain.
The dependence of the interface mobility on temperature is crucial
to obtain this behavior. As described in our previous paper \cite{bayle2020}, the mobility is obtained from
measurements of the crystallization velocity of pure GST as a function of temperature \cite{orava2012}. As the temperature decreases from the melting temperature, the velocity
first increases because of the growing thermodynamic driving force, but then decreases again
because all atomic rearrangement processes slow down. Therefore, when, as in Fig.~\ref{fig_reset_structure}, the temperature decreases very rapidly, the interface motion
cannot keep up with the temperature field, and the liquid is quenched into the solid amorphous state.
The final structure is consistent with experimental measurements \cite{sousa2015}.

\subsection{SET operation and germanium redistribution}
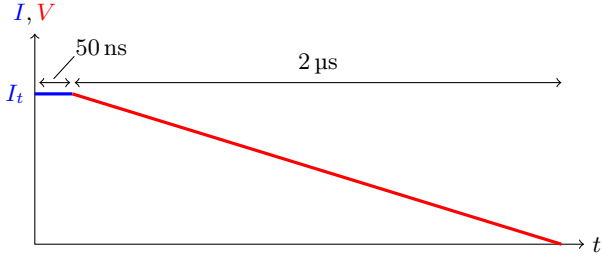
\begin{figure}[!ht]
	\centering
	\begin{tikzpicture}
		\def\tUp{0.5}						
		\pgfmathsetmacro\tDown{13 * \tUp}	
		\pgfmathsetmacro\xMax{\tDown + \tUp + 0.3}		
		\def\yMax{2.8}		
		\def\vPulse{2.0}	

		\draw [<->] (0, \yMax) node [above] {${\color{blue}I}, {\color{red}V}$}
			-- (0, 0) -- (\xMax, 0) node [right] {$t$};

		\draw [blue, very thick] (0,    \vPulse) node [left] {$I_t$} -- (\tUp, \vPulse);
		\draw [red,  very thick] (\tUp, \vPulse) -- (\tUp + \tDown, 0);

		\def\hFl{0.15}	
		\def\eFl{0.06}	
		\draw [<->] (\eFl,          \hFl + \vPulse) -- (\tUp - \eFl/2, \hFl + \vPulse);
		\draw (\tUp/2, \hFl + \vPulse + 0.05) -- (\tUp*1.05, \hFl + \vPulse + 0.35)
			node [above right=-0.1cm] {\qty{50}{\nano\second}};
		\draw [<->] (\eFl/2 + \tUp, \hFl + \vPulse) -- (\tUp + \tDown, \hFl + \vPulse)
			node [pos=.5, above] {$\qty{2}{\micro\second}$};
	\end{tikzpicture}
	\caption{Electrical pulse used for SET simulations.
	The initial plateau lasts \qty{50}{\nano\second}, like in RESET operations.
	Two colors are used to separate the plateau and the ramp down: the former is controlled with current and the latter is controlled with the voltage.}
	\label{fig_pulse_SET}
\end{figure}

\begin{figure}[!ht]
	\centering
	\includegraphics[width=0.34\textwidth]{legende_microstructure.png} \vspace{0.12cm}\\
	\includegraphics[width=0.47\textwidth]{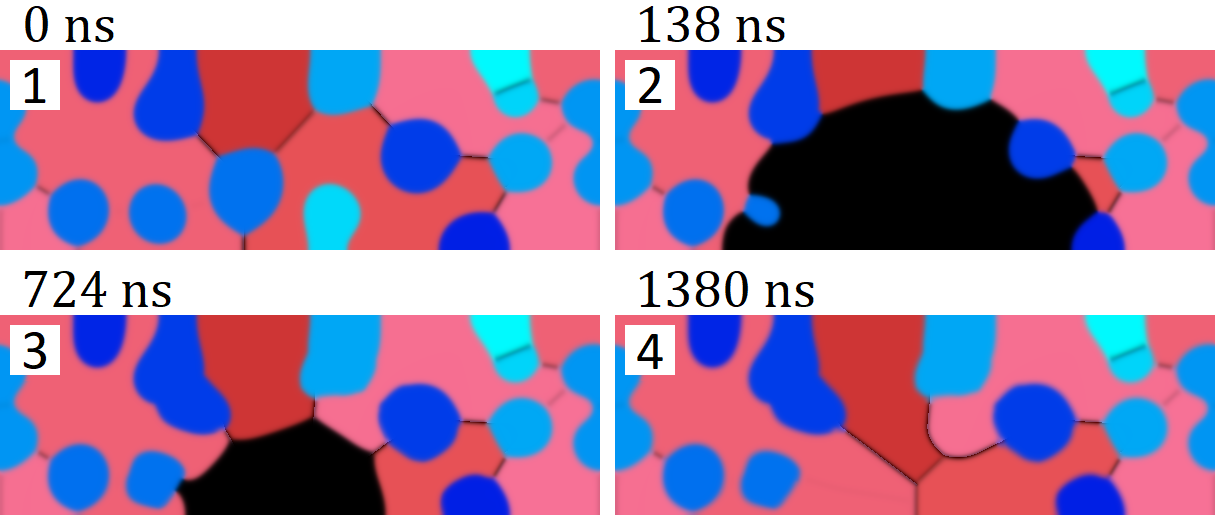}
	\caption{Microstructure evolution during a SET pulse: from the largest dome after initial heating (2, roughly after \qty{135}{\nano\second}) to a fully crystallized layer (4).}
	\label{fig_set_structure}
\end{figure}

The programming pulse used to simulate a SET operation is presented in Fig. \ref{fig_pulse_SET}.
Due to the progressive decrease of the imposed current, the evolution of the microstructure is different (see Fig. \ref{fig_set_structure}).
A similar melting occurs, but the temperature of the crystal/liquid interface stays high enough during cooldown to enable full recrystallization of the active material.

\begin{figure}[!ht]
	\centering
	\subfloat[Disks positions relatively to the final microstructure.]{
		\centering
		\begin{tikzpicture} [every node/.style={inner sep=0,outer sep=0}]
			\node[anchor=south west] at (0,0) {
				\includegraphics[width=0.37\textwidth]{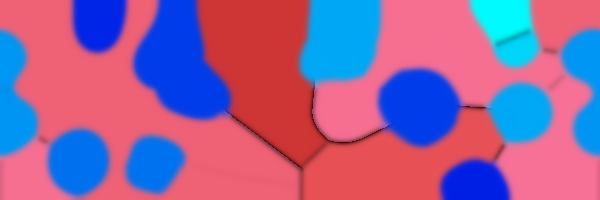}};
			\path [fill=bluyan, very thick] (5.50, 0) arc [radius=2.20, start angle=0, end angle= 180] -- (5.50, 0);
			\path [fill=vert,   very thick] (4.95, 0) arc [radius=1.65, start angle=0, end angle= 180] -- (4.95, 0);
			\path [fill=orange, very thick] (4.40, 0) arc [radius=1.10, start angle=0, end angle= 180] -- (4.40, 0) ;
		\end{tikzpicture}
		\label{fig_SET_disk}
	}
	\par
	\subfloat[Concentration $c$ at the end of SET operation.]{
		\centering
		\begin{tikzpicture} [every node/.style={inner sep=0,outer sep=0}]
			\node[anchor=south west] at (0,0) {
				\includegraphics[width=0.385\textwidth]{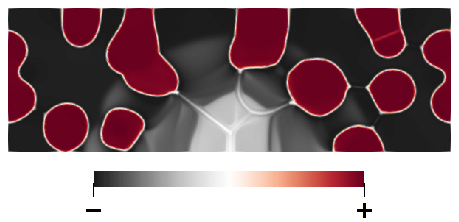}};
			\draw [bluyan, very thick] (5.58, 1.05) arc [radius=2.16, start angle=0, end angle= 180];
			\draw [vert, very thick] (5.04, 1.05) arc [radius=1.62, start angle=0, end angle= 180];
			\draw [orange, very thick] (4.50, 1.05) arc [radius=1.08, start angle=0, end angle= 180];
		\end{tikzpicture}
		\label{fig_SET_c}
	}
	\caption{Positions of the annuli and concentration map at the end of the SET operation presented in Fig. \ref{fig_set_structure} (4th image).
	}
	\label{fig_SET_disk_c}
\end{figure}
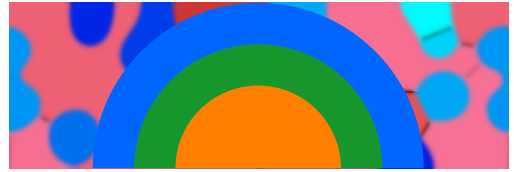
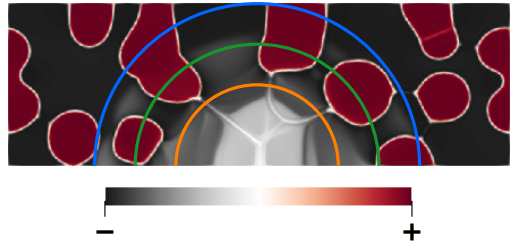

\begin{figure*}[t]
	\centering
	\includegraphics[width=0.77\textwidth]{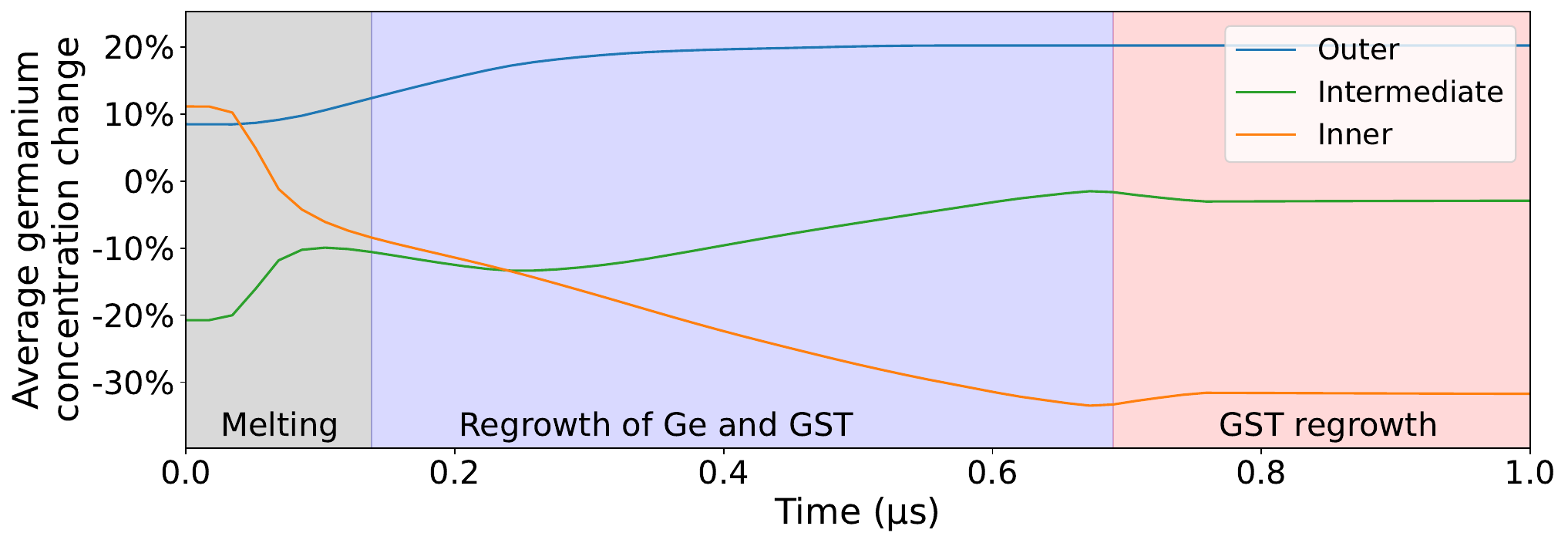}
	\caption{Evolution of the average value of $c$ (with respect to the initial concentration $c_0$) on the annuli presented in Fig. \ref{fig_SET_disk_c} (same colors are used).
	The time axis is divided into three parts corresponding to the three phases of the SET operation: melting, regrowth of both germanium and GST, regrowth of GST only.}
	\label{fig_SET_evol_c_disk}
\end{figure*}

Comparing the snapshots 1 and 4 of Fig. \ref{fig_set_structure}, we can 
see that all the germanium grains that were located at the center of the melted zone have 
disappeared, whereas some germanium grains that were at the edge of the melted dome
end up being larger than before the operation. Thus, there has been a net germanium
redistribution from the center to the border of the melted dome.
This is also observed experimentally \cite{sousa2015}.
To characterize this 
effect more precisely, we consider three annuli with increasing radius (see Fig. \ref{fig_SET_disk_c}) and we track the evolution of the average excess germanium concentration during the SET operation, which is displayed in Fig. \ref{fig_SET_evol_c_disk}.
The time axis is divided into three parts corresponding to the three phases of the operation.
During the initial melting of the material, the outer annulus (only partially melted) sees its concentration slightly increase, while the two other areas (fully melted) homogenize.
After that, during the regrowth of the material, as the crystal/liquid front moves toward the center of the memory cell, first, the concentration of the outer annulus increases, and then the one of the intermediate annulus.
At the same time, the concentration of the liquid (green and orange curves first, then only orange) steadily decreases.
This is due to two factors: to grow, germanium grains must absorb germanium from the liquid, and, moreover, the concentration of regrown GST is higher than at the beginning (see the lighter shade of gray on Fig. \ref{fig_SET_c}).
The last phase of regrowth starts slightly before the snapshot picture of
Fig. \ref{fig_set_structure}.3, when the growth of the germanium grains stops.
Since, in the further evolution, only GST recrystallizes, germanium segregates into the 
liquid, whose concentration increases in consequence.
This explains why, in Fig. \ref{fig_SET_c}, the closer to the heater (where the last liquid area recrystallized), the higher the concentration.
Fig. \ref{fig_SET_evol_c_disk} confirms that after the operation, a germanium redistribution toward the edge of the melted dome occurred.
The localized accumulation of germanium in the liquid at the end of the recrystallization process is of second order to this global redistribution.

The model is able to reproduce the trend observed experimentally: the germanium concentration increases at the edge of the melted dome while the concentration of the core is lower than initially.
However, the magnitude of the simulated concentration redistribution is lower than in experiments.
Moreover, the large increase of the liquid concentration at the end of recrystallization is not
seen in the experiments. More detailed comparisons would be necessary to determine the exact 
origins of these discrepancies.

\subsection{Electrothermal fields in the segregated material} \label{ss_result_electrothermal}
\begin{figure}[!ht]
	\centering
	\subfloat[Temperature (in kelvin), low current.]{
		\centering
		\includegraphics[width=0.382\textwidth]{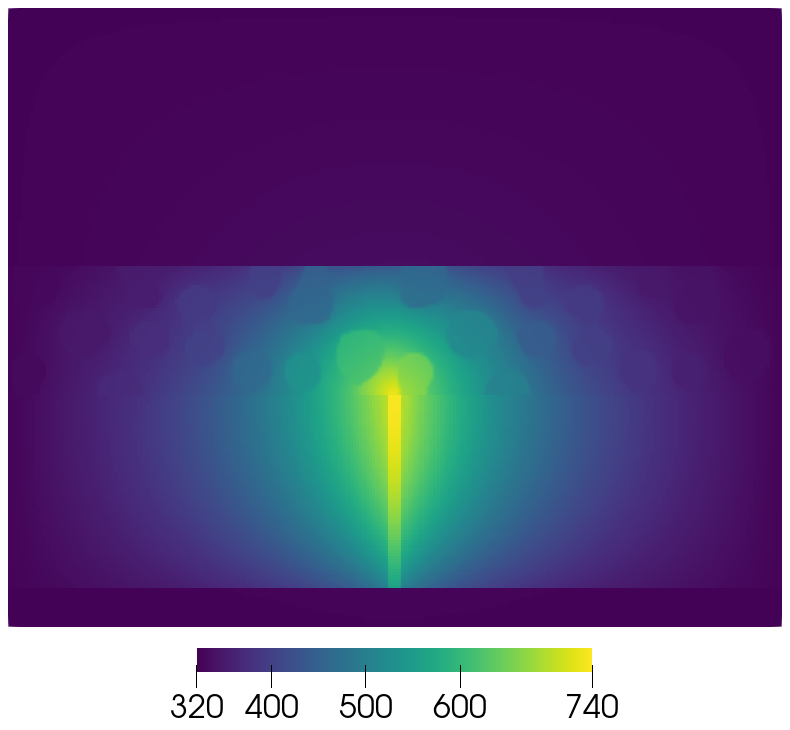}
		\label{fig_temp_low_current}
	}
	\par
	\subfloat[Temperature (in kelvin), high current.]{
		\centering
		\includegraphics[width=0.382\textwidth]{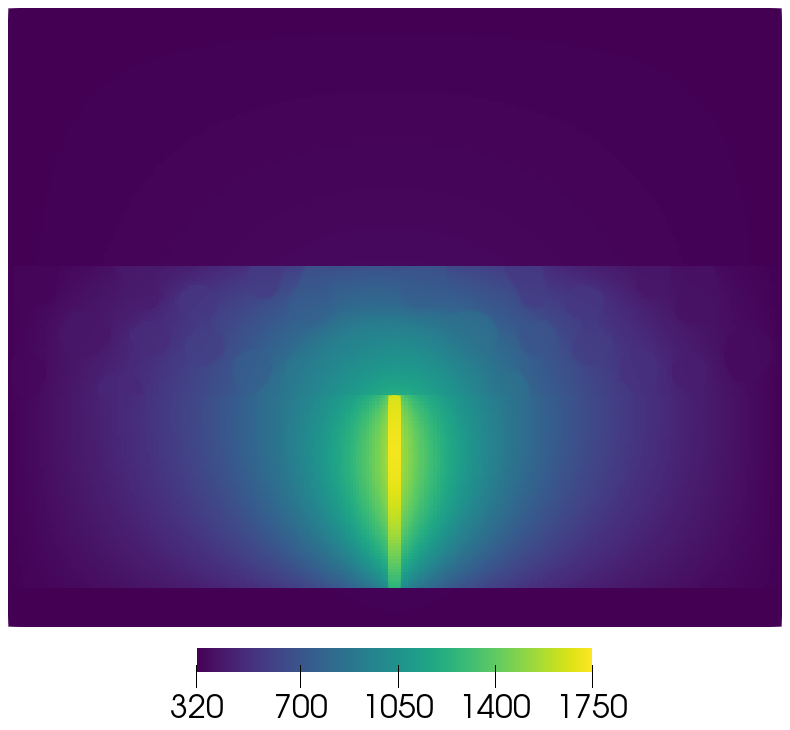}
		\label{fig_temp_high_current}
	}
	\caption{Temperature in the full memory domain, during two RESET simulations using different current, after \qty{50}{\nano\second}.
	The corresponding microstructure are respectively Fig. \ref{fig_initial_microstructure} and Fig. \ref{fig_reset_structure}.3.}
	\label{fig_temp_low_high_current}
\end{figure}

The coupled models can be exploited to visualize the electrothermal fields during memory operations, in the GGST segregated microstructure.
First, we can look at the temperature profile in the memory.
Fig. \ref{fig_temp_low_high_current} shows the temperature in the full domain at \qty{50}{\nano\second}, in two situations: at the top, the programming current is too low to melt the material (Fig. \ref{fig_initial_microstructure} is the corresponding microstructure);  at the bottom, the current is higher (same simulation as Fig. \ref{fig_reset_structure}).
In Fig. \ref{fig_temp_low_current}, the large differences in thermal properties of germanium and GST are visible: germanium grains clearly stand out.
In Fig. \ref{fig_temp_high_current}, the temperature is more homogeneous in the melted dome.
Another important difference is that if the material stays crystalline, the temperature is almost equal at the top of the heater and at the bottom of GGST, whereas if it melts, the temperature drops in the phase change material.
These two figures also highlight the impact of TBR: a sharp drop of several hundred kelvins occurs at the interfaces of the memory element, in line with values obtained in the literature \cite{durai2020, reifenberg2006}. \vbl

\begin{figure}[!ht]
	\centering
	\subfloat[Current density (in \unit{\ampere\per\meter\squared}).]{
		\centering
		\includegraphics[width=0.38\textwidth]{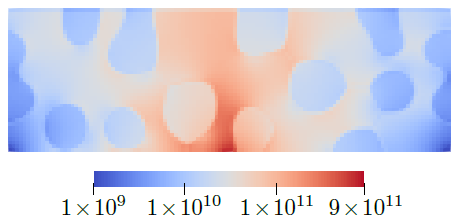}
		\label{fig_j_cryst}
	}
	\par
	\subfloat[Joule heating (in \unit{\watt\per\meter\cubed}).]{
		\centering
		\includegraphics[width=0.38\textwidth]{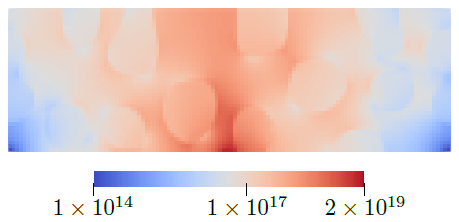}
		\label{fig_joule_cryst}
	}
	\caption{Current density and Joule heating in GGST before melting (microstructure of Fig. \ref{fig_initial_microstructure}).}
\end{figure}

\begin{figure}[!ht]
	\centering
	\includegraphics[width=0.38\textwidth]{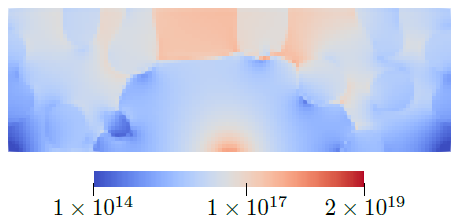}
	\caption{Joule heating (in \unit{\watt\per\meter\cubed}) after the melting of a dome in the microstucture (see Fig. \ref{fig_reset_structure}.3).}
	\label{fig_joule_melt}
\end{figure}

We can also look at spatial maps of current density and Joule heating.
In Fig. \ref{fig_j_cryst}, it can be seen that the electric current avoids germanium grains because of their higher electrical resistivity.
This leads to a reduced Joule heating in these grains (see Fig. \ref{fig_joule_cryst}).
On Fig. \ref{fig_joule_melt}, after the melting of the material, the Joule heating has drastically decreased due to the high electrical conductivity of the liquid phase.

To avoid any confusion, one must understand that unlike the two temperature maps of Fig. \ref{fig_temp_low_high_current} that correspond to different programming current (but are both obtained at the end of a \qty{50}{\nano\second} RESET pulse), Fig. \ref{fig_joule_cryst} and \ref{fig_joule_melt} corresponds to the same programming current.
The former is obtained close to the beginning of the pulse (after less than \qty{1}{\nano\second}) whereas the latter is obtained after \qty{50}{\nano\second}.

\subsection{Electrical figures of merit}
\begin{figure}[!ht]
	\centering
	\includegraphics[width=0.44\textwidth]{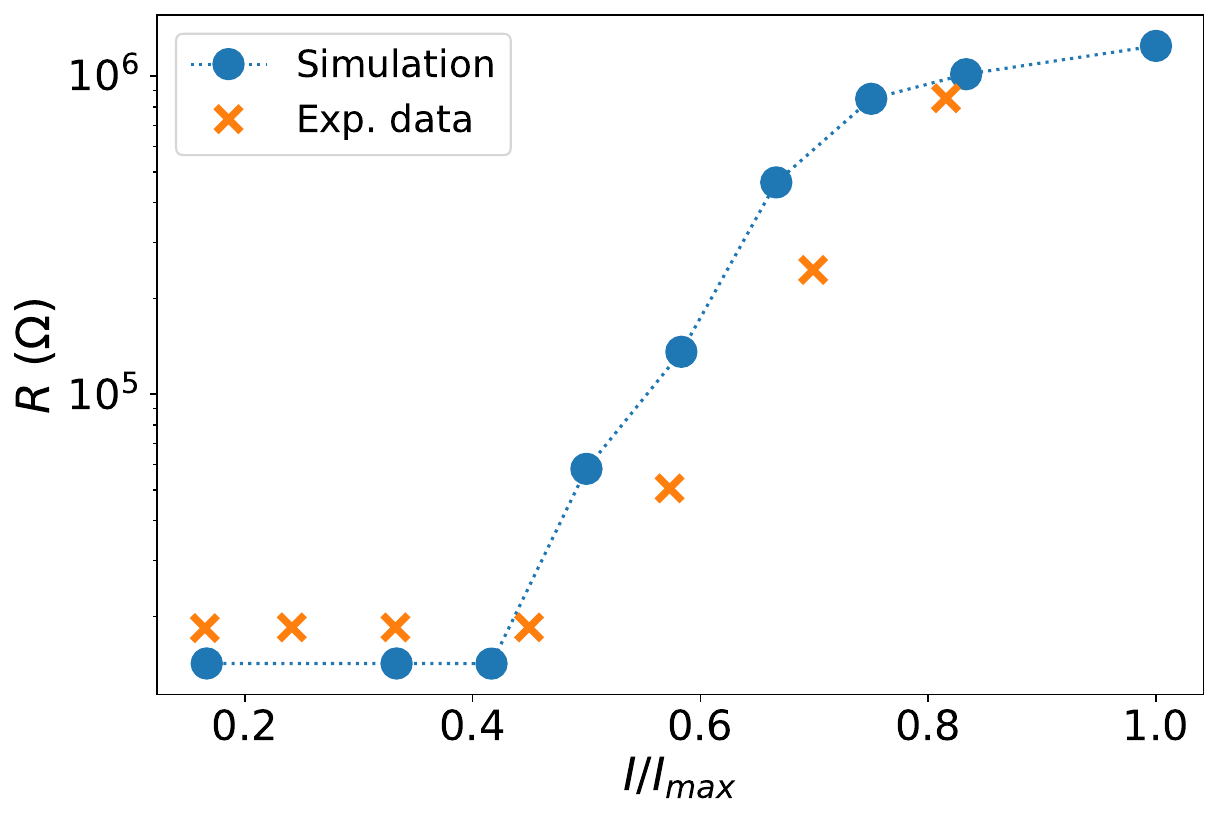}
	\caption{Comparison between simulated and experimental $R(I)$ curve.
	Normalized electric current $I/I_{\max}$ used.}
	\label{fig_R(I)}
\end{figure}

\begin{figure}[!ht]
	\centering
	\includegraphics[width=0.44\textwidth]{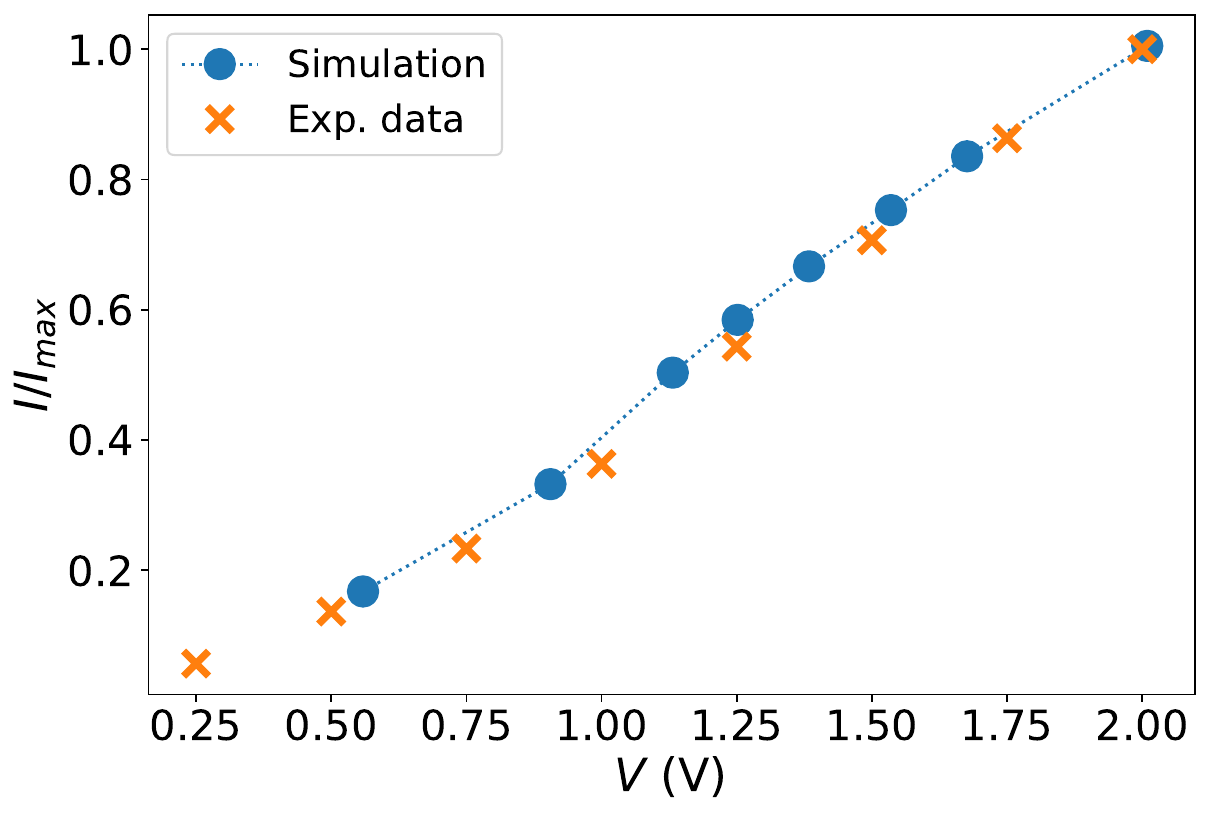}
	\caption{Comparison between simulated and experimental $I(V)$ curve.
	Normalized electric current $I/I_{\max}$ used.}
	\label{fig_I(V)}
\end{figure}

\begin{figure}[!ht]
	\centering

	\def\yMax{2.8}		
	\def\vPulse{2.0}	
	\def\tUp{1.2}							
	\pgfmathsetmacro\tDown{\tUp * 3/5}		
	\def\xMax{3.2}		

	\subfloat[$R(I)$]{
		\centering
		\begin{tikzpicture}
			\draw [<->] (0, \yMax) node [above] {$I$} -- (0, 0) -- (\xMax, 0) node [right] {$t$};

			\draw [blue, very thick] (0,    \vPulse) node [left] {$I_t$}
			                       -- (\tUp, \vPulse) -- (\tUp, 0) -- (\tUp + \tDown, 0);

			\def\eFl{0.1}	
			\draw [decorate, decoration = {brace, raise=3pt, amplitude=5pt}]
				(\eFl, \vPulse) -- (\tUp + \tDown, \vPulse) node [midway, above=8pt] {RESET};

			\draw [<-] (\tUp + \tDown, 0.1) -- (\tUp + \tDown, 0.6) node [above] {READ};
		\end{tikzpicture}
	}
	\subfloat[$I(V)$]{
		\centering
		\begin{tikzpicture}
			\draw [<->] (0, \yMax) node [above] {$I$} -- (0, 0) -- (\xMax, 0) node [right] {$t$};

			\draw [blue, very thick] (0, \vPulse) node [left] {$I_t$} -- (\tUp*2, \vPulse);

			\def\hFl{0.15}	
			\def\eFl{0.06}	
			\draw [<->] (\eFl, \hFl + \vPulse) -- (\tUp*2 - \eFl, \hFl + \vPulse)
				node [pos=.5, above] {\qty{100}{\nano\second}};

			\draw [<-] (\tUp*2, \vPulse - 0.1) -- (\tUp*2, \vPulse - 0.6)
				node [below] {$V$ and $I$ read};
		\end{tikzpicture}
	}
	\caption{Electrical pulses for $R(I)$ and $I(V)$ curves.
	For each data point of Fig. \ref{fig_R(I)} and \ref{fig_I(V)}, a simulation with a different value of $I_t$ is needed.
	Notice how in (b), $V$ and $I$ are read while the current is still equal to $I_t$.}
	\label{fig_RI_IV_pulses}
\end{figure}
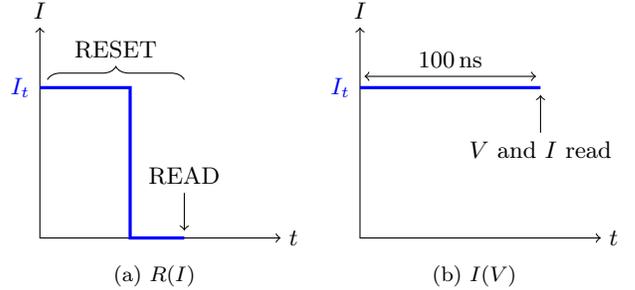

We have performed simulations to reproduce two figures of merit that are commonly used 
to characterize the memory devices: a $R(I)$ plot in Fig. \ref{fig_R(I)} 
and a $I(V)$ plot in Fig. \ref{fig_I(V)}.
The $R(I)$ curve is obtained by performing several RESET operations at different programming current
intensities $I$, followed by the reading of $R$, the total resistance of the memory cell.
Each data point corresponds to one simulation.
Similarly, the $I(V)$ curve is obtained by performing several simulations with different $I$.
Those simulations are longer RESET pulses (\qty{100}{\nano\second} instead of \qty{50}{\nano\second} as shown in Fig. \ref{fig_pulse_RESET}), which lead to an almost stabilized microstructure.
The voltage $V$ across the cell is read before the end of the pulse.
The result of this procedure is $V$ as a function of $I$, which can then be inverted to yield the more customary current-voltage characteristic $I(V)$ for comparison with data from experiments.
The two pulses are illustrated in Fig. \ref{fig_RI_IV_pulses}.

In Fig. \ref{fig_R(I)}, the model reproduces experimental measurements well.
At a certain current, the resistance starts to increase because the Joule heating is high enough to melt (and, after the cooling, amorphize) a part of the active material.
The effect saturates at higher currents since the dome radius reaches the GGST layer thickness.
The resistances at low and high currents correspond, respectively, to the values of the 
fully crystallized layer and of the amorphous phase.
The agreement between simulated and experimental $I(V)$ curves in Fig. \ref{fig_I(V)} is also excellent.
At high current, when the active material melts and becomes significantly more conductive, the drop of the applied voltage essentially occurs in the heater.
This can be used to determine the electrical properties of this material (see Sec.~\ref{ss_calib_heater}).\vbl

\begin{figure}[!ht]
	\centering
	\begin{minipage}[b]{0.23\textwidth}
		\centering
		\includegraphics[width=\columnwidth]{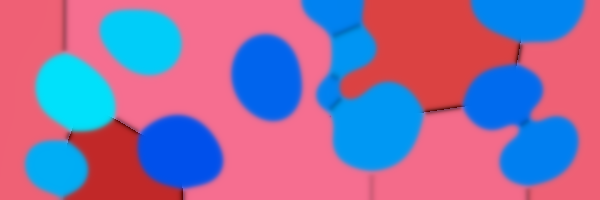}
	\end{minipage}
	\hspace{0.02cm}
	\begin{minipage}[b]{0.23\textwidth}
		\centering
		\includegraphics[width=\columnwidth]{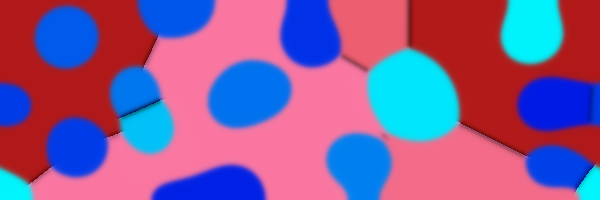}
	\end{minipage}
	\vspace{0.2cm}\\
	\begin{minipage}[b]{0.23\textwidth}
		\centering
		\includegraphics[width=\columnwidth]{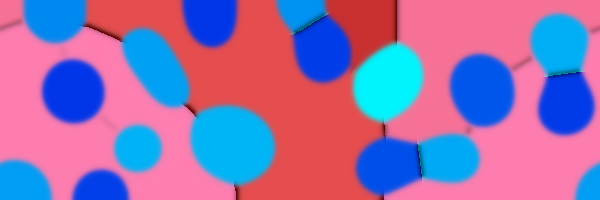}
	\end{minipage}
	\hspace{0.02cm}
	\begin{minipage}[b]{0.23\textwidth}
		\centering
		\includegraphics[width=\columnwidth]{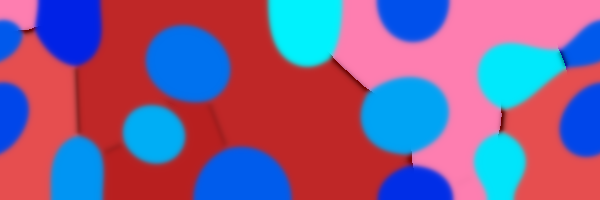}
	\end{minipage}
	\vspace{0.2cm}\\
	\begin{minipage}[b]{0.23\textwidth}
		\centering
		\includegraphics[width=\columnwidth]{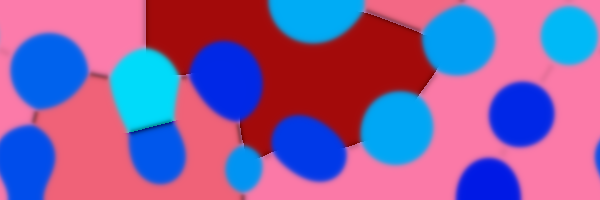}
	\end{minipage}
	\hspace{0.02cm}
	\begin{minipage}[b]{0.23\textwidth}
		\centering
		\includegraphics[width=\columnwidth]{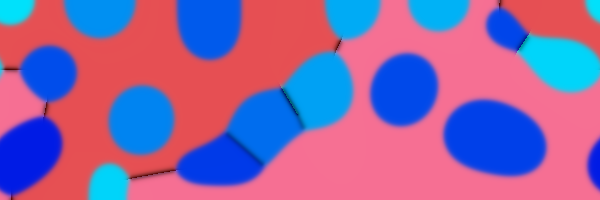}
	\end{minipage}
	\vspace{0.2cm}\\
	\begin{minipage}[b]{0.23\textwidth}
		\centering
		\includegraphics[width=\columnwidth]{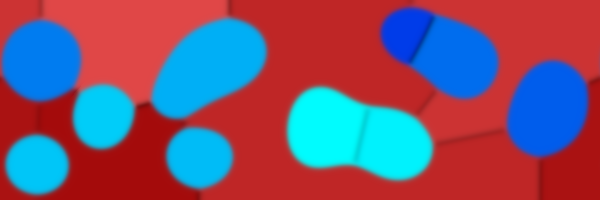}
	\end{minipage}
	\hspace{0.02cm}
	\begin{minipage}[b]{0.23\textwidth}
		\centering
		\includegraphics[width=\columnwidth]{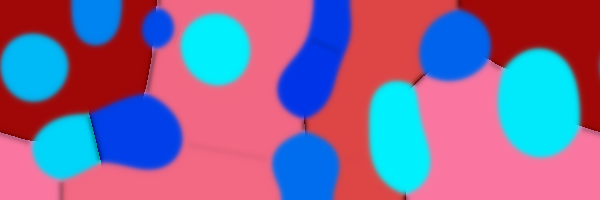}
	\end{minipage}
	\vspace{0.2cm}\\
	\begin{minipage}[b]{0.23\textwidth}
		\centering
		\includegraphics[width=\columnwidth]{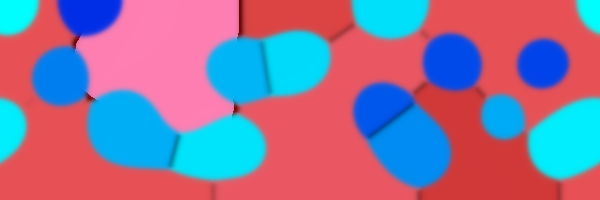}
	\end{minipage}
	\caption{Additional initial microstructures.}
	\label{fig_RI_stat_microstructure}
\end{figure}

The impact of the initial microstructure has also been studied.
As explained in Section \ref{ss_initial_cond}, all the memory operation simulations presented 
so far have used Fig. \ref{fig_initial_microstructure} as a starting point.
However, variations in the microstructure, in particular the initial position of the germanium grains, can lead to different electrical behavior of the memory.
A total of nine additional microstructures were generated randomly by multiple annealing simulations (see Fig. \ref{fig_RI_stat_microstructure}), and both $R(I)$ and $I(V)$ simulations were performed with each of them.

\begin{figure}[!ht]
	\centering
	\subfloat[$R(I)$ simulations]{
		\centering
		\includegraphics[width=0.44\textwidth]{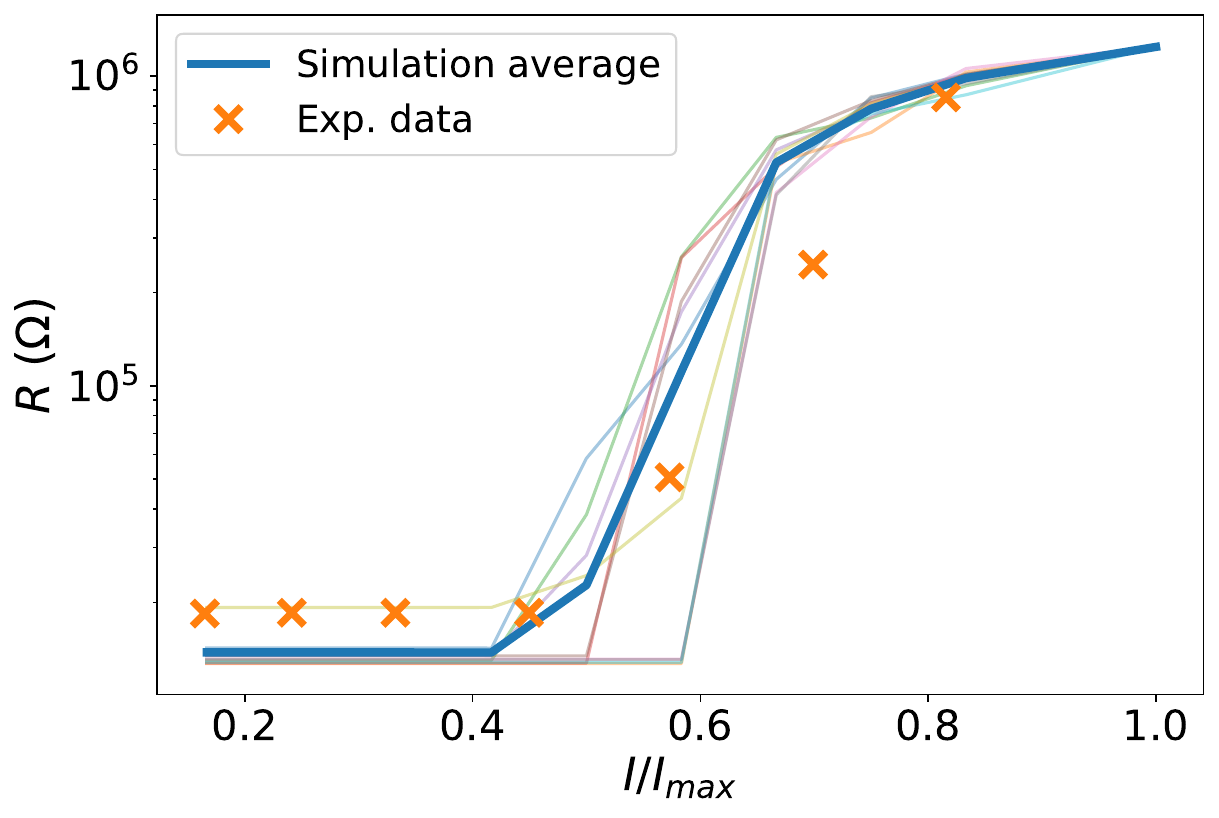}
		\label{fig_RI_stat}
	}
	\par
	\subfloat[$I(V)$ simulations]{
		\centering
		\includegraphics[width=0.44\textwidth]{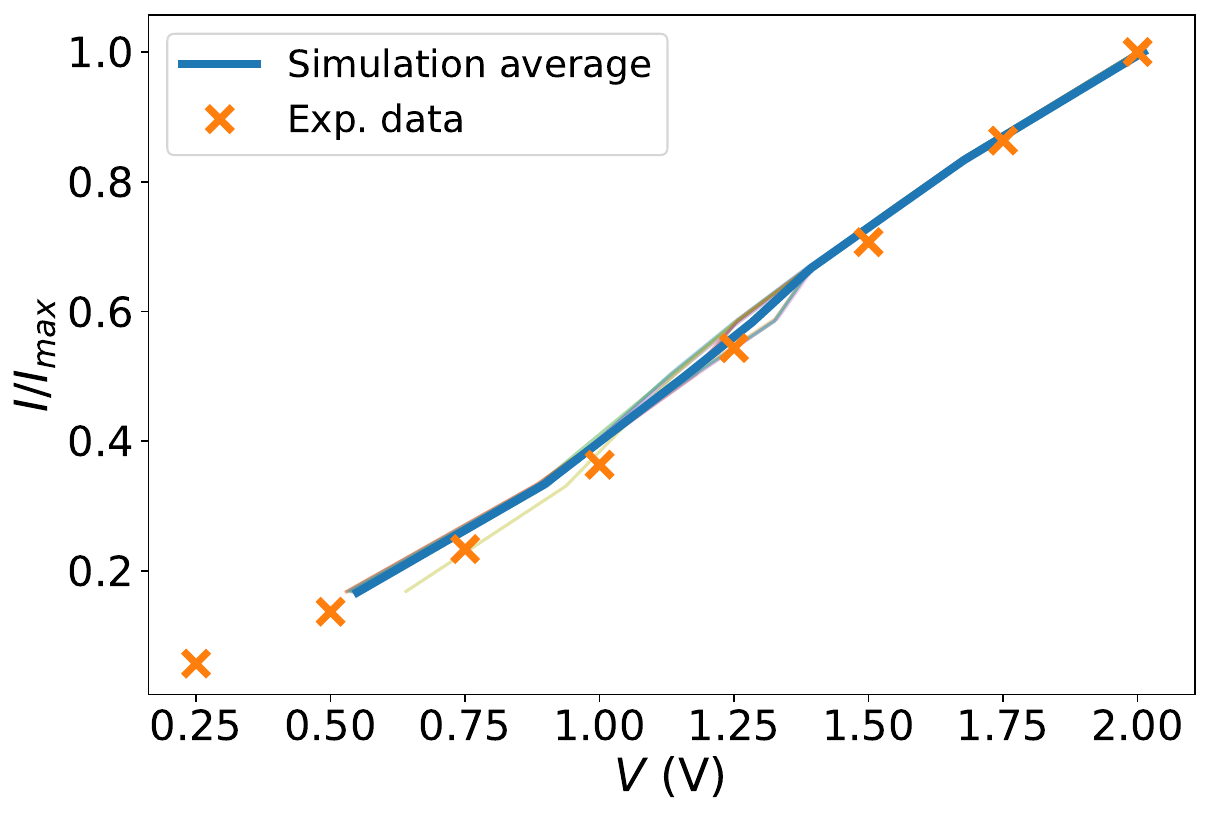}
		\label{fig_IV_stat}
	}
	\caption{Average of $R(I)$ and $I(V)$ simulations performed on various initial microstructures.
	Individual simulations are displayed in the background with thinner lines.
	Same normalization as in Fig. \ref{fig_R(I)} and \ref{fig_I(V)}.}
	\label{fig_stat}
\end{figure}

In Fig. \ref{fig_stat}, the curves associated with the different microstructures are averaged and compared to experimental data.
Individual curves are also displayed with thinner lines to visualize the variability of the results.
On the $R(I)$ plot, individual curves are scattered; in particular, the onset current leading to 
the formation of the melted dome varies significantly.
However, when taking the average of the ten curves, a good agreement with experimental data emerges.

\subsection{Model calibration: discussion} \label{ss_calibration}
Now, we will discuss the process that has allowed us to calibrate the model parameters
and to obtain the good agreement with experiments that was displayed above.

\subsubsection{Thermal conductivity of crystalline Ge-rich GST} \label{ss_calib_kthGGST}
The thermal conductivity of crystalline GGST, that is, the two-phase polycrystalline
material, has been measured up to \qty{700}{\kelvin} \cite{kusiak2022}. With
the values of the thermal conductivities gathered from the literature, simulations
with our model, using the microstructure of Fig. \ref{fig_initial_microstructure}, 
yielded an effective thermal conductivity (that is, the ratio of
total heat flux through the entire sample, divided by the temperature difference
between top and bottom) that was far too large. This problem was cured in two steps.

First, it was realized that the thermal conductivity of germanium thin films \cite{wang2011} is significantly lower than the one initially used in our model, corresponding to bulk germanium samples \cite{glassbrenner1964}.
Unfortunately, the literature does not provide thermal conductivity values for germanium thin films over the whole 300-\qty{1200}{\kelvin} temperature range.
Instead, it was decided to divide the values reported for bulk germanium samples by a factor of 8; this division is already included in the data shown in Fig.~\ref{fig_kth_3_phases}.
Even with this modification, the model still yields an effective thermal conductivity 
that is more than \qty{60}{\percent} higher than the value measured at \qty{300}{\kelvin}; to obtain the experimental
value, a reduction of the germanium conductivity by a factor of 40 would have been
necessary, which does not seem reasonable. Instead, TBR are added to the model at 
germanium/GST interfaces. Indeed, just as the ``device'' interfaces between dissimilar
materials, the ``internal'' interfaces inside the active material should add thermal
resistance.

To include those ``internal'' TBR, two points must be taken into account.
First, contrary to ``device'' interfaces, germanium/GST interfaces can move during a 
simulation, and have to be located at any time. For this purpose, the product $g_1\,g_2$ is used.
Indeed, outside of the 1--2 interface (the germanium/GST interface), either $g_1$ or $g_2$ is zero ($p_1\,p_2$ could also be used but $g_1\,g_2$ is more localized).
Second, interfaces in the crystallization model (in the MPFM) are diffuse and span across multiple nodes of the fine computational grid, which means that the TBR is also ``smeared out'' over several grid points.
Noting $S_g^{12}$ the total integral of the $g_1\,g_2$ product in the direction normal to the interface (depends on the MPFM's parameters, $S_g^{12} = 0.7$ in our case), the thermal conductivity in the GGST
becomes
\begin{equation}
\kth = \cfrac{1}{ \cfrac{1}{ \kth^{mat} } + \cfrac{g_1 \, g_2}{S_g^{12}} \cfrac{ R_{12} }{ dx } }
\end{equation}
with $\kth^{mat}$ the bulk thermal conductivity (see Eq. \ref{eq_kth123}), and $R_{12}$ the TBR of the interface.
With this expression, the thermal conductivity of nodes out of a germanium/GST interface remains equal to $\kth^{mat}$.

\begin{figure}[!ht]
	\centering
	\includegraphics[width=0.44\textwidth]{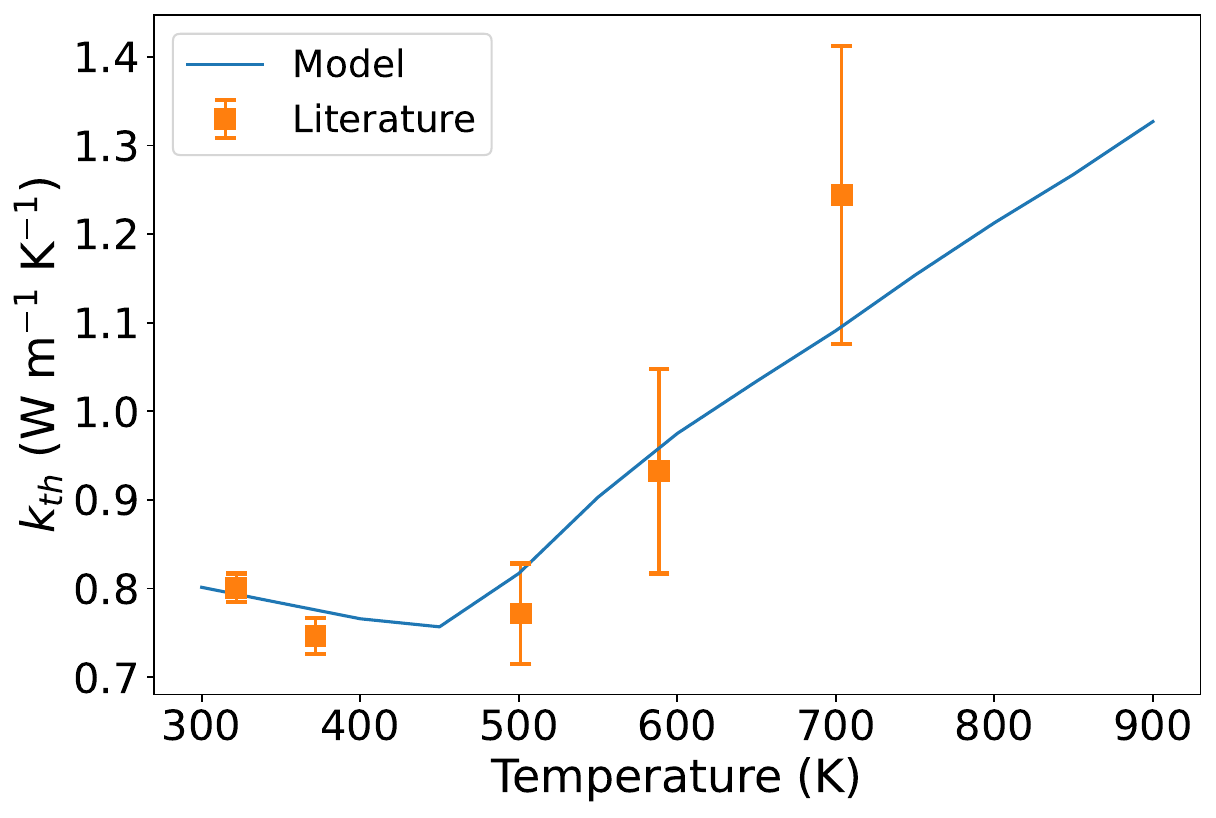}
	\caption{Simulated thermal conductivity of crystalline GGST when TBR are included at germanium/GST interfaces.
	Literature data come from [\onlinecite{kusiak2022}].}
	\label{fig_kth_GGST_TBR}
\end{figure}

The value of $R_{12}$ has been calibrated to yield the correct value of the effective thermal conductivity of the two-phase polycrystalline material at \qty{300}{\kelvin} (\qty{0.8}{\kth}, as in [\onlinecite{kusiak2022}]).
As shown in Fig. \ref{fig_kth_GGST_TBR}, simulations with this value of TBR match well with experimental measurement over an extended temperature range.
Furthermore, the value $R_{12} = \qty{8e-9}{\kelvin\meter\squared\per\watt}$ is of the right order of magnitude when compared to the values for the other interfaces provided in Table \ref{tab_TBR}.
In summary, including TBR in the GGST microstructure seems to be important to accurately reproduce the thermal behavior of the material.

\subsubsection{Resistivity of the heater} \label{ss_calib_heater}
As mentioned along Fig. \ref{fig_I(V)}, the electrical properties of the heater can be determined from the $I(V)$ curve.
Initially, the electrical conductivity of the material was set at \qty{6.5e4}{\sig}, according to in-house measurements done at low temperatures (up to a few hundred degrees Celsius).
However, with this value, the simulated current was systematically too high.
Since the temperature is much higher during memory operations (especially for 
high currents), an increase of the electrical resistivity can be expected, as in most metals.
Good agreement with the experimental $I(V)$ curve was obtained with the value 
of \qty{5.0e4}{\sig} already given in Sec.~\ref{ss_elconductivity}.

\subsubsection{Melting of germanium grains during RESET} \label{ss_calib_fonteGe}
With the initial parameters, we observed that most germanium grains located
inside the liquid dome did not completely melt during a RESET pulse. Indeed, since the
melting temperature of GST is lower than the one of germanium, the melting process
was faster in the GST matrix, and isolated germanium grains remained within the liquid.
After cooling, thus, crystalline germanium inclusions were present in the amorphized dome. 
This was never observed in the experiments. We had initially thought that the inclusion 
of the phase-dependent conductivity would resolve this issue, since the higher resistivity
of germanium should lead to locally stronger Joule heating. However, this is offset by
the inhomogeneous current distribution shown in Fig.~\ref{fig_j_cryst}. Therefore, the
melting of the germanium grains must be accomplished by heat generated in the heater and transported
through the liquid. To promote this melting process, we have lowered the thermal coductivity
of the liquid by a factor of 5; this leads to higher temperatures close to the heater 
inside the liquid, since the heat is more strongly confined. The data shown in 
Fig.~\ref{fig_kth_3_phases} do not include this factor.

\subsubsection{Progressive development of the amorphous dome}  \label{ss_calib_RI}
The last adjustments concern the $R(I)$ curve (Fig.~\ref{fig_R(I)}).
Initially, the transition between small and large values of $R(I)$ was very abrupt,
even if at the same time the size of the amorphous dome was increasing gradually with the current.
To obtain the more progressive behavior shown in the figure, it was necessary to
take into account the Poole-Frenkel effect as given in Sec.~\ref{ss_elconductivity}.
Indeed, this effect reduces the resistivity of small amorphous domes, which leads
to a more progressive increase of resistance with the size of the amorphous dome.

On this aspect, the electrical conductivity of the liquid is also key.
Indeed, thanks to its high value, there is almost no Joule heating in the melted dome as shown in Fig. \ref{fig_joule_melt}.
This enables more progressive dome sizes by removing most of the Joule heating in the GGST as the melted area expends.

\section{Conclusion}
In order to model the evolution of the GGST microstructure during memory operations, a fully self-consistent coupling of the MPFM with a thermal model and an electrical model has been presented.
The thermal model accounts for heat generation and its propagation in the memory, while the electrical model provides the electric current density (used to compute the Joule heating) and is used to determine the electrical resistance of the memory.
The numerous additional parameters needed have been obtained from literature data and in-house material characterization. The complete model has then been calibrated to match the experimental data.

SET and RESET operation simulations demonstrate the ability of this model to qualitatively reproduce the GGST amorphization and crystallization depending on the programming conditions.
It is also possible to exploit the model to study the evolution of electro-thermal fields (temperature, current density...) during operations.
For instance, we showed that the melting of germanium grains cannot be explained by a higher Joule heating (compared to GST) due to their high electrical resistivity.
Beyond that, our simulations of electrical figures of merit are in good agreement with experimental data, even when considering changes in the microstructure of the material.
Finally, we have also evidenced that thermal boundary resistances were needed at germanium/GST interfaces to properly reproduce the thermal conductivity measured on crystalline GGST.

Despite these good results, some caveats are in order, and some lines for future improvement
can be outlined. Most importantly, our simulations are two-dimensional. The real three-dimensioal
grain structure is very different from the columnar structure that would be obtained if our 
two-dimensional heat maps are extended along the third dimension. Furthermore, the geometry of the
conduction paths inside the GST phase would be quite different in two and three dimensions. However,
it is known that grain coarsening exhibits the same scaling laws in two and three dimensios, only with
different prefactors. Since the low-conductivity germanium phase forms isolated islands and does never percolate, there should be a quantitative, but no qualitative difference concerning conduction.
This explains why, although two-dimensional, our model still captures many relevant features.
For simulations in three dimensions, the orientation-field model needs to be adapted. Indeed, in
three dimensions the orientation of a grain needs to be specified by three scalar parameters, which
can be the classic Euler angles or the elements of a unit quaternion \cite{Pusztai05}. Furthermore, 
keeping acceptable simulation times despite the large number of discretization points would 
require more sophisticated parallelization methods, such as MPI or GPU methods. This is an 
interesting subject for future work. As an additional benefit, such three-dimensional models 
could potentially capture the full scope of three-dimensional complexities inherent to many 
real-world device architectures with arbitrary boundary shapes.
 
Another important point is the lack of knowledge about parameters. We had to determine many of them
approximately due to the lack of more precise information in the literature; this mainly applies to 
the liquid phase, but not exclusively. A systematic sensitivity analysis could reveal which of these
parameters are most influential for the device performance. However, in view of the large number
of parameters, this represents an enormous amount of work, which is outside the scope of the present article. 
As an alternative, advanced simulation methods such as molecular dynamics with neural network-trained interatomic potentials could be exploited \cite{kheir2024} to obtain more precise values at least for some parameters.

Another area of improvement is the crystallization model.
In this paper, we mainly focused of the thermal and electrical models and on their calibration, however, some aspects of the crystallization model, such as its kinetics, play a key role in the physics of the device.
To this end, simulation results could for instance be compared to experimental measurement of the SET resistance as a function of the quenching time.
Also, the pseudobinary approximation limits the composition of the system (see Sec. \ref{ss_model_cryst}).
In particular, it is not possible to include pure Sb lamellas as reported recently in the literature \cite{rahier2023}.
However, extending the current multi-phase field model to the full ternary phase diagram is very challenging.
Set aside the complex mathematical formulation, it would require a lot of additional physical parameters, notably to model the thermodynamic behavior of all the different phases.

Currently, this model already provides an excellent framework to investigate diverse programming conditions and material change. 
In particular, it is able to simulate the forming operation, the first electrical activation needed by GGST alloy in order to exhibit good switching behavior.
This forming operation is known to deeply influence the microstructure \cite{palumbo2017}.
While the results presented in this paper have not considered post-forming microstructure as starting point for SET and RESET operations, future exploitation of the models could 
reveal how these operations are impacted by the forming step.
Moreover, the capability of the models to explore various programming pulses extends the models utility to address the multistate capability of the PCM.
This capacity of multistate is of first importance for in-memory computing \cite{sebastian2019} and neuromorphic applications, where the gradual change of resistance is instrumental 
in mimicking the analog nature of biological synapses.

\begin{acknowledgments}
This work was financially supported by the Association Nationale Recherche Technologie
(ANRT), France, and STMicroelectronics through the CIFRE contract number 2020/1070.
\end{acknowledgments}

\section*{Data Availability}
The data that support the findings of this study, and in particular the simulation
code, are available from the corresponding author upon reasonable request.

\appendix
\section{Crystallization model equations}
In this appendix, all the equations of the crystallization model that are used in the simulations are summarized for completeness; full details and further explanations can be found in our previous paper \cite{bayle2020}.

\subsection{Multi-phase field model}
Denoting $\mathbf{p} = (p_1,p_2,p_3)$, the grand potential functional (the energy functional) used to describe the system is
\begin{align}
\begin{split}
	\Omega(\textbf{p},\nabla \textbf{p}, T, \mu)
	= \int_V \bigg(
			& \frac{K}{2}\omega_\text{grad}(\nabla \textbf{p})
			  + H\omega_{tw}(\textbf{p}) \\
			& + \omega_{th}(\textbf{p}, T, \mu) \bigg) dV,
\end{split}
\end{align}
with $K$ and $H$ two parameters that set the width of the diffuse interfaces $W=\sqrt{K/H}$ 
and its surface energy, and
\begin{align}
	\omega_\text{grad}(\nabla \textbf{p}) 	&= \sum_{i=1}^3 (\nabla p_i)^2, \\
	\omega_{tw} (\textbf{p}) 				&= \sum_{i=1}^3 p_i^2(1-p_i)^2, \\
	\omega_{th} (\textbf{p}, T, \mu) 		&= \sum_{i=1}^3 g_i(\textbf{p})\omega_i(T,\mu). \label{eq_omegaTH}
\end{align}
The functions $\omega_i(T,\mu)$ are the grand potentials for each phase, obtained by a
Legendre transform of the free energy $f(T,c)$, where $c$ is a composition variable and
$\mu$ its thermodynamically conjugate chemical potential.

The interpolation functions $g_i$ (one for each phase) that appear in Eq. (\ref{eq_omegaTH}), 
but also in Eq.~(\ref{eq_kth123}) or Eq.~(\ref{eq_source}) are given by
\begin{align}
\begin{split}
	g_i(\textbf{p}) &= g(p_i, p_j, p_k) \\[0.1cm]
	                &= \frac{p_i^2}{4}\Big(
							15(1-p_i)\left[1 + p_i - (p_j-p_k)^2\right] \\[0.1cm]
						& \qquad \quad + p_i(9p_i^2 - 5)
						\Big),
\end{split}
\end{align}
their sum is equal to 1 (as for $p_i$).

The time evolution of the three phase fields $p_i$ are obtain by solving the following relaxation equation (taking into account the constraint on the sum $p_1 + p_2 + p_3 = 1$):
\begin{equation}
	\tau(T) \fpartial{p_i}{t} =
		-\frac{1}{H} \left.\!\frac{\delta \Omega}{\delta p_i}
		\right|_{\sum p\, = 1},
	\label{eq_evol_pi}
\end{equation}
by deriving the functional, we obtain:
\begin{align}
\begin{split}
	\tau&(T) \fpartial{p_i}{t} =
		-\frac{K}{H} \nabla^2 p_i \\[0.1cm]
		&-\frac{2}{3} \bigg( \!
			-2p_i(1-p_i)(1-2p_i) + \sum_{\phi = j,k} p_\phi(1-p_\phi)(1-2p_\phi)
			\bigg) \\[-0.1cm]
		&-\frac{1}{H} \sum_{\phi=i,j,k} \omega_\phi(T, \mu)
			\left. \! \fpartial{g_\phi(\textbf{p})}{p_i} \right|_{\sum p \, = 1}
\end{split}
\end{align}

\subsection{Chemical diffusion}
To take into account chemical diffusion, a mass conservation law together with
a transport law of linear irreversible thermodynamics are used:
\begin{equation}
	\frac{\partial c}{\partial t}
		= \vv{\nabla} \cdot \left[
			M(\textbf{p}, T, \mu) \vv{\nabla} \mu(\textbf{p}, T, c)
		\right],
	\label{2_eq_evol_c}
\end{equation}
with $\mu$ the chemical potential and $M$ the atomic mobility. We work in a mixed
formulation, in which both $c$ and $\mu$ are used as variables for different purposes:
the diffusion currents are evaluated with the help of $\mu$, whereas mass conservation
is treated in $c$. Both variables are linked in the grand-canonical formalism by
\begin{equation}
	c = \sum_{i = 1}^3 g_i(\textbf{p}) c_i(T, \mu),
	\label{2_eq_c_mu}
\end{equation}
with $c_i(T, \mu)=-\partial\omega_i/\partial\mu$.
Depending on $\textbf{p}$ and $T$, we ``search'', by dichotomy, the value of $\mu$ correponding to the concentration found with Eq. \ref{2_eq_evol_c}.

\subsection{Orientation field model}
The orientation field model modifies Eq. \ref{eq_evol_pi} for the two crystalline phases ($i = 1,2$), an additional term is considered:
\begin{equation}
	\tau(T)\fpartial{p_i}{t} =
		-\frac{1}{H} \left.\!\frac{\delta \Omega}{\delta p_i}
		\right|_{\sum p\, = 1}
		- \frac{C_\theta}{H} q'(p_i) (\nabla \theta_i)^2,
	\label{2_eq_evol_p_orientation}
\end{equation}
with $C_\theta$ that determine the strength of coupling between the phase field and the orientation field.
The function $q(p)$ verifies
\begin{equation}
	q(p) = \frac{7p^3 - 6p^4}{(1-p)^3}.
\end{equation}
The evolution equation for the two orientation fields $\theta_i$ is
\begin{equation}
	\tau_{\theta_i}(T) \fpartial{\theta_i}{t} =
		\frac{C_\theta}{Hq(p_i)} \vv{\nabla} \cdot \left( q(p_i) \vv{\nabla} \theta_i \right).
	\label{2_eq_evol_orientation}
\end{equation}
Because of this expression, $p_1$ and $p_2$ cannot reach the values 0 and 1.

\section{Crystallization model parameters}
\subsection{Thermodynamics}
\begin{figure}[!ht]
	\centering
	\includegraphics[width=0.44\textwidth]{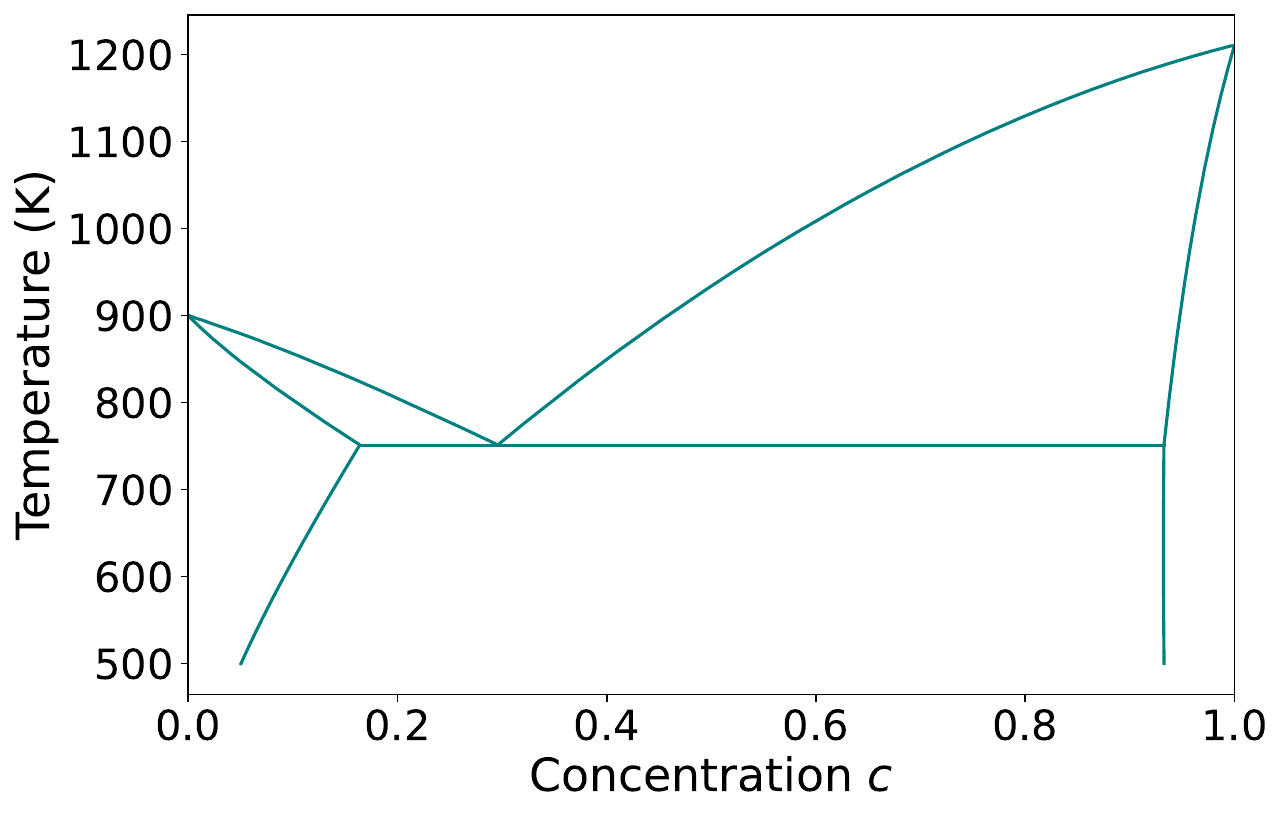}
	\caption{Phase diagram.}
	\label{fig_diagram}
\end{figure}

For the definition of the free-energy and grand potential functions for each phase,
we use a regular solution model. We use the same functions as [\onlinecite{bayle2020}], with
the same nomenclature, but with slightly different values for some of the coefficients;
all the parameters used to generate the phase diagram are listed in Table \ref{tab_diag_param},
and the phase diagram is displayed in Fig.~\ref{fig_diagram}.
\begin{table}[!ht]
	\centering
	\caption{Phase diagram parameters.}
	\label{tab_diag_param}

	\begin{tabular}{@{} llrl @{}}
	\toprule
									& Symbol			& Value				& Unit			\\
	\midrule
	Mixing coefficients				& $\Omega_{GST}$ 	& \num{-1.50e4}		& \unit{\joule\per\mole} \\
									& $\Omega_{Ge}$ 	& \num{-1.24e5}		& \unit{\joule\per\mole} \\
									& $\Omega_{liq}$ 	& \num{-1.0e4}		& \unit{\joule\per\mole} \\
	Latent heats of fusion 			& $L_{GST}$  		& \num{1.2e4}		& \unit{\joule\per\mole} \\
									& $L_{Ge}$  		& \num{3.7e4}		& \unit{\joule\per\mole} \\
	Melting temperatures  			& $T_m^{GST}$ 		& \num{900}			& \unit{\kelvin}		\\
									& $T_m^{Ge}$		& \num{1211}		& \unit{\kelvin} 		\\
									& $T_m^X$ 			& \num{243.75}		& \unit{\kelvin}		\\
									& $T_m^Y$ 			& \num{-3201.21}	& \unit{\kelvin} 		\\
	Equilibrium concentrations	 	& $c_{GST,eq}$  	& \num{0.10}		&						\\
									& $c_{Ge,eq}$  		& \num{0.95}		&						\\
	Partition coefficients			& $k_{GST}$ 		& \num{0.3125}		&						\\
									& $k_{Ge}$ 			& \num{0.0735}		&						\\
	Eutectic temperature  			& $T_e$  			& \num{800}			& \unit{\kelvin}		\\
	Eutectic concentration 			& $c_e$ 			& \num{0.32}		&						\\
	\bottomrule
	\end{tabular}
\end{table}

\subsection{Other parameters}
Other constant parameters of the crystallization model are listed in Table \ref{tab_autres_param}.
\begin{table}[!ht]
	\centering
	\caption{Crystallization model parameters.}
	\label{tab_autres_param}

	\begin{tabular}{@{} llrl @{}}
	\toprule
								& Symbol			& Value				& Unit			\\
	\midrule
	\qquad\; ---				& $K$ 				& \num{6.90e-15}	& \unit{\joule\meter\squared\per\meter} \\
	\qquad\; ---				& $H$ 				& \num{2.76e4}		& \unit{\joule\per\mole} \\
	Interface width 			& $W$ 				& \num{0.5e-9}		& \unit{\meter} \\
	Interface surface energy	& $\gamma$  		& \num{0.4}			& \unit{\joule\per\meter\squared} \\
	\qquad\; ---				& $C_{\theta}/H$  	& \num{8.33e-26}	& \unit{\meter\squared} \\
	Molar volume				& $V_m$				& \num{1.63e-5}		& \unit{\meter\cubed\per\mole} \\
	\bottomrule
	\end{tabular}
\end{table}
The temperature dependence of the kinetic coefficient $\tau$ is the same as in [\onlinecite{bayle2020}] and
has been calibrated on measured data for pure GST \cite{orava2012}.
The two kinetic coefficients of the orientation field model are defined with respect to $\tau(T)$:
\begin{align}
	\tau_{\theta_1}(T) &\simeq \num{8.33e-8}\tau(T), \\
	\tau_{\theta_2}(T) &= 5 \, \tau_{\theta_1}(T).
\end{align}

Finally, the mobility $M$ is linked to the diffusivity $D$ by $M=D\chi$, where
$\chi=\partial^2\omega/\partial\mu^2$ is the susceptibility function. The expression 
for the diffusivity is interpolated similarly to the thermal and electrical conductivities:
\begin{equation}
	D(\textbf{p}, T) = g_1 D_1(T) + g_2 D_2(T) + g_3 D_3(T),
	\label{2_eq_diffusivite}
\end{equation}
with $D_i$ given by
\begin{equation}
	D_i = D_0\exp\left(-\frac{\alpha}{k_BT}\right),
\end{equation}
with coefficients given in Table \ref{tab_D_coef}.
\begin{table}[!ht]
	\centering
	\caption{Coefficient of the diffusivities $D_i$.}
	\label{tab_D_coef}
	\begin{tabular}{@{} llcc @{}} \toprule
							& & $D_0$				& $\alpha$ (\unit{\joule})	\\
	\midrule
	Germanium and GST $(i = 1,2)$ 	& & \num{2.40e-9}	& \num{8.10e-20}	\\
	Amorphous and liquid $(i = 3)$	& & \num{1.86e-8} 	& \num{3.46e-20}	\\
	\bottomrule
	\end{tabular}
\end{table}


\begin{thebibliography}{10}

\bibitem{redaelli2022}
A.~Redaelli, E.~Petroni, and R.~Annunziata, ``Material and process engineering
  challenges in {Ge}-rich {GST} for embedded {PCM},'' {\em Materials Science in
  Semiconductor Processing}, vol.~137, p.~106184, Jan. 2022.

\bibitem{yamada1991}
N.~Yamada, E.~Ohno, K.~Nishiuchi, N.~Akahira, and M.~Takao, ``Rapid phase
  transitions of {GeTe}‐{Sb2Te3} pseudobinary amorphous thin films for an
  optical disk memory,'' {\em Journal of Applied Physics}, vol.~69,
  pp.~2849--2856, Mar. 1991.

\bibitem{zuliani2013}
{P. Zuliani}, {E. Varesi}, {E. Palumbo}, {M. Borghi}, {I. Tortorelli}, {D.
  Erbetta}, {G. D. Libera}, {N. Pessina}, {A. Gandolfo}, {C. Prelini}, {L.
  Ravazzi}, and {R. Annunziata}, ``Overcoming {Temperature} {Limitations} in
  {Phase} {Change} {Memories} {With} {Optimized} \ce{Ge_xSb_yTe_z},'' {\em IEEE
  Transactions on Electron Devices}, vol.~60, pp.~4020--4026, Dec. 2013.

\bibitem{luong2021}
M.~A. Luong, M.~Agati, N.~Ratel~Ramond, J.~Grisolia, Y.~Le~Friec, D.~Benoit,
  and A.~Claverie, ``On {Some} {Unique} {Specificities} of {Ge}-{Rich} {GeSbTe}
  {Phase}-{Change} {Material} {Alloys} for {Nonvolatile} {Embedded}-{Memory}
  {Applications},'' {\em physica status solidi (RRL) – Rapid Research
  Letters}, vol.~15, p.~2000471, Mar. 2021.

\bibitem{bayle2020}
R.~Bayle, O.~Cueto, S.~Blonkowski, T.~Philippe, H.~Henry, and M.~Plapp,
  ``Phase-field modeling of the non-congruent crystallization of a ternary
  {Ge}–{Sb}–{Te} alloy for phase-change memory applications,'' {\em Journal
  of Applied Physics}, vol.~128, p.~185101, Nov. 2020.

\bibitem{durai2020}
{S. Durai}, {S. Raj}, and {A. Manivannan}, ``Impact of {Thermal} {Boundary}
  {Resistance} on the {Performance} and {Scaling} of {Phase}-{Change} {Memory}
  {Device},'' {\em IEEE Transactions on Computer-Aided Design of Integrated
  Circuits and Systems}, vol.~39, pp.~1834--1840, Sept. 2020.

\bibitem{miquel2023}
{R. Miquel}, {T. Cabout}, {O. Cueto}, {B. Sklénard}, and {M. Plapp}, ``A
  {Fully} {Coupled} {Multi}-{Physics} {Model} to {Simulate} {Phase} {Change}
  {Memory} {Operations} in {Ge}-rich {Ge2Sb2Te5} {Alloys},'' in {\em 2023
  {International} {Conference} on {Simulation} of {Semiconductor} {Processes}
  and {Devices} ({SISPAD})}, pp.~317--320, Sept. 2023.

\bibitem{sousa2015}
{V. Sousa}, {G. Navarro}, {N. Castellani}, {M. Coué}, {O. Cueto}, {C.
  Sabbione}, {P. Noé}, {L. Perniola}, {S. Blonkowski}, {P. Zuliani}, and {R.
  Annunziata}, ``Operation fundamentals in {12Mb} {Phase} {Change} {Memory}
  based on innovative {Ge}-rich {GST} materials featuring high reliability
  performance,'' in {\em 2015 {Symposium} on {VLSI} {Technology} ({VLSI}
  {Technology})}, pp.~T98--T99, June 2015.

\bibitem{kato2005}
T.~Kato and K.~Tanaka, ``Electronic {Properties} of {Amorphous} and
  {Crystalline} {Ge2Sb2Te5} {Films},'' {\em Japanese Journal of Applied
  Physics}, vol.~44, p.~7340, Oct. 2005.

\bibitem{baldo2020}
{M. Baldo}, {O. Melnic}, {M. Scuderi}, {G. Nicotra}, {M. Borghi}, {E. Petroni},
  {A. Motta}, {P. Zuliani}, {A. Redaelli}, and {D. Ielmini}, ``Modeling of
  virgin state and forming operation in embedded phase change memory ({PCM}),''
  in {\em 2020 {IEEE} {International} {Electron} {Devices} {Meeting} ({IEDM})},
  pp.~13.3.1--13.3.4, Dec. 2020.

\bibitem{pirovano2004}
{A. Pirovano}, {A. L. Lacaita}, {F. Pellizzer}, {S. A. Kostylev}, {A.
  Benvenuti}, and {R. Bez}, ``Low-field amorphous state resistance and
  threshold voltage drift in chalcogenide materials,'' {\em IEEE Transactions
  on Electron Devices}, vol.~51, pp.~714--719, May 2004.

\bibitem{glassbrenner1964}
C.~J. Glassbrenner and G.~A. Slack, ``Thermal {Conductivity} of {Silicon} and
  {Germanium} from {3K} to the {Melting} {Point},'' {\em Physical Review},
  vol.~134, pp.~A1058--A1069, May 1964.

\bibitem{wang2011}
Z.~H. Wang and M.~J. Ni, ``Thermal conductivity and its anisotropy research of
  germanium thin film,'' {\em Heat and Mass Transfer}, vol.~47, pp.~449--455,
  Apr. 2011.

\bibitem{lyeo2006}
H.-K. Lyeo, D.~G. Cahill, B.-S. Lee, J.~R. Abelson, M.-H. Kwon, K.-B. Kim,
  S.~G. Bishop, and B.-k. Cheong, ``Thermal conductivity of phase-change
  material {Ge2Sb2Te5},'' {\em Applied Physics Letters}, vol.~89, p.~151904,
  Oct. 2006.

\bibitem{agati2019}
M.~Agati, M.~Vallet, S.~Joulié, D.~Benoit, and A.~Claverie, ``Chemical phase
  segregation during the crystallization of {Ge}-rich {GeSbTe} alloys,'' {\em
  Journal of Materials Chemistry C}, vol.~7, no.~28, pp.~8720--8729, 2019.

\bibitem{crespi2014}
{L. Crespi}, {A. Ghetti}, {M. Boniardi}, and {A. L. Lacaita}, ``Electrical
  {Conductivity} {Discontinuity} at {Melt} in {Phase} {Change} {Memory},'' {\em
  IEEE Electron Device Letters}, vol.~35, pp.~747--749, July 2014.

\bibitem{kusiak2022}
A.~Kusiak, C.~Chassain, A.~M. Canseco, K.~Ghosh, M.-C. Cyrille, A.~L. Serra,
  G.~Navarro, M.~Bernard, N.-P. Tran, and J.-L. Battaglia,
  ``Temperature-{Dependent} {Thermal} {Conductivity} and {Interfacial}
  {Resistance} of {Ge}-{Rich} {Ge2Sb2Te5} {Films} and {Multilayers},'' {\em
  physica status solidi (RRL) – Rapid Research Letters}, vol.~16, p.~2100507,
  Apr. 2022.

\bibitem{baratella2022}
D.~Baratella, D.~Dragoni, and M.~Bernasconi, ``First-{Principles} {Calculation}
  of {Transport} and {Thermoelectric} {Coefficients} in {Liquid} {Ge2Sb2Te5},''
  {\em physica status solidi (RRL) – Rapid Research Letters}, vol.~n/a,
  p.~2100470, Sept. 2022.

\bibitem{assael2017}
M.~J. Assael, K.~D. Antoniadis, W.~A. Wakeham, M.~L. Huber, and H.~Fukuyama,
  ``Reference {Correlations} for the {Thermal} {Conductivity} of {Liquid}
  {Bismuth}, {Cobalt}, {Germanium}, and {Silicon},'' {\em Journal of Physical
  and Chemical Reference Data}, vol.~46, p.~033101, Sept. 2017.

\bibitem{poling2001book}
B.~E. Poling, J.~M. Prausnitz, and J.~P. O’Connell, {\em Properties of
  {Gases} and {Liquids}}.
\newblock New York: McGraw-Hill Education, fifth edition.~ed., 2001.

\bibitem{smith1966}
R.~C. Smith, ``High-{Temperature} {Specific} {Heat} of {Germanium},'' {\em
  Journal of Applied Physics}, vol.~37, pp.~4860--4865, Dec. 1966.

\bibitem{chen1969}
H.~S. Chen and D.~Turnbull, ``Specific {Heat} and {Heat} of {Crystallization}
  of {Amorphous} {Germanium},'' {\em Journal of Applied Physics}, vol.~40,
  pp.~4214--4215, Sept. 1969.

\bibitem{kalb2002Thesis}
J.~Kalb, {\em Stresses, viscous flow and crystallization kinetics in thin films
  of amorphous chalcogenides used for optical data storage}.
\newblock PhD thesis, Aachen University, Apr. 2002.

\bibitem{scott2020}
E.~A. Scott, E.~Ziade, C.~B. Saltonstall, A.~E. McDonald, M.~A. Rodriguez,
  P.~E. Hopkins, T.~E. Beechem, and D.~P. Adams, ``Thermal conductivity of
  ({Ge2Sb2Te5})1-{xCx} phase change films,'' {\em Journal of Applied Physics},
  vol.~128, p.~155106, Oct. 2020.

\bibitem{taylor1964}
R.~E. Taylor and J.~Morreale, ``Thermal {Conductivity} of {Titanium} {Carbide},
  {Zirconium} {Carbide}, and {Titanium} {Nitride} at {High} {Temperatures},''
  {\em Journal of the American Ceramic Society}, vol.~47, pp.~69--73, Feb.
  1964.

\bibitem{mohammadpour2018}
E.~Mohammadpour, M.~Altarawneh, J.~Al-Nu’airat, Z.-T. Jiang, N.~Mondinos, and
  B.~Z. Dlugogorski, ``Thermo-mechanical properties of cubic titanium
  nitride,'' {\em Molecular Simulation}, vol.~44, pp.~415--423, Mar. 2018.

\bibitem{white1997}
G.~K. White and M.~L. Minges, ``Thermophysical properties of some key solids:
  {An} update,'' {\em International Journal of Thermophysics}, vol.~18,
  pp.~1269--1327, Sept. 1997.

\bibitem{serra2019}
{A. L. Serra}, {O. Cueto}, {N. Castellani}, {J. Sandrini}, {G. Bourgeois}, {N.
  Bernier}, {M. C. Cyrille}, {J. Garrione}, {M. Bernard}, {V. Beugin}, {A.
  André}, {J. Guerrero}, {G. Navarro}, and {E. Nowak}, ``Outstanding
  {Improvement} in {4Kb} {Phase}-{Change} {Memory} of {Programming} and
  {Retention} {Performances} by {Enhanced} {Thermal} {Confinement},'' in {\em
  2019 {IEEE} 11th {International} {Memory} {Workshop} ({IMW})}, pp.~1--4, May
  2019.

\bibitem{NIST_JANAF1998book}
{M Chase}, {\em {NIST}-{JANAF} {Thermochemical} {Tables}}.
\newblock American Institute of Physics, 4th~ed., Aug. 1998.

\bibitem{ranica2021}
R.~Ranica, R.~Berthelon, A.~Gandolfo, G.~Samanni, E.~Gomiero, J.~Jasse,
  P.~Mattavelli, J.~Sandrini, M.~Querre, Y.~Le-Friec, J.~Poulet, V.~Caubet,
  L.~Favennec, C.~Boccaccio, G.~Ghezzi, C.~Gallon, J.~Grenier, B.~Dumont,
  O.~Weber, A.~Villaret, R.~Beneyton, N.~Cherault, D.~Ristoiu, S.~Del~Medico,
  O.~Kermarrec, J.~Reynard, P.~Boivin, A.~Souhaite, L.~Desvoivres, S.~Chouteau,
  P.~Sassoulas, L.~Clement, A.~Valery, E.~Petroni, D.~Turgis, A.~Lippiello,
  L.~Scotti, F.~Disegni, A.~Ventre, D.~Ornaghi, M.~De~Tomasi, A.~Maurelli,
  A.~Conte, F.~Arnaud, A.~Redaelli, R.~Annunziata, P.~Cappelletti, F.~Piazza,
  P.~Ferreira, R.~Gonella, and E.~Ciantar, ``Heater system optimization for
  robust epcm reliability and scalability in 28nm fdsoi technology,'' in {\em
  2021 IEEE International Electron Devices Meeting (IEDM)}, pp.~28.1.1--28.1.4,
  2021.

\bibitem{samani2013}
M.~Samani, X.~Ding, S.~Amini, N.~Khosravian, J.~Cheong, G.~Chen, and B.~Tay,
  ``Thermal conductivity of titanium aluminum silicon nitride coatings
  deposited by lateral rotating cathode arc,'' {\em Thin Solid Films},
  vol.~537, pp.~108--112, June 2013.

\bibitem{lee2013}
J.~Lee, E.~Bozorg-Grayeli, S.~Kim, M.~Asheghi, H.-S. Philip~Wong, and K.~E.
  Goodson, ``Phonon and electron transport through {Ge2Sb2Te5} films and
  interfaces bounded by metals,'' {\em Applied Physics Letters}, vol.~102,
  p.~191911, May 2013.

\bibitem{reifenberg2010}
{J. P. Reifenberg}, {K. Chang}, {M. A. Panzer}, {S. Kim}, {J. A. Rowlette}, {M.
  Asheghi}, {H. -. P. Wong}, and {K. E. Goodson}, ``Thermal {Boundary}
  {Resistance} {Measurements} for {Phase}-{Change} {Memory} {Devices},'' {\em
  IEEE Electron Device Letters}, vol.~31, pp.~56--58, Jan. 2010.

\bibitem{battaglia2010}
J.-L. Battaglia, A.~Kusiak, V.~Schick, A.~Cappella, C.~Wiemer, M.~Longo, and
  E.~Varesi, ``Thermal characterization of the {SiO2}-{Ge2Sb2Te5} interface
  from room temperature up to 400°{C},'' {\em Journal of Applied Physics},
  vol.~107, p.~044314, Feb. 2010.

\bibitem{jeong2012}
T.~Jeong, J.-G. Zhu, S.~Chung, and M.~R. Gibbons, ``Thermal boundary resistance
  for gold and {CoFe} alloy on silicon nitride films,'' {\em Journal of Applied
  Physics}, vol.~111, p.~083510, Apr. 2012.

\bibitem{kittel2004book}
C.~Kittel, {\em Introduction to {Solid} {State} {Physics}}.
\newblock Wiley, 8th~ed., Nov. 2004.

\bibitem{jiang2002}
J.~Z. Jiang, H.~Lindelov, L.~Gerward, K.~Ståhl, J.~M. Recio, P.~Mori-Sanchez,
  S.~Carlson, M.~Mezouar, E.~Dooryhee, A.~Fitch, and D.~J. Frost,
  ``Compressibility and thermal expansion of cubic silicon nitride,'' {\em
  Physical Review B}, vol.~65, p.~161202, Apr. 2002.

\bibitem{ciocchini2014}
{N. Ciocchini}, {E. Palumbo}, {M. Borghi}, {P. Zuliani}, {R. Annunziata}, and
  {D. Ielmini}, ``Modeling {Resistance} {Instabilities} of {Set} and {Reset}
  {States} in {Phase} {Change} {Memory} {With} {Ge}-{Rich} {GeSbTe},'' {\em
  IEEE Transactions on Electron Devices}, vol.~61, pp.~2136--2144, June 2014.

\bibitem{eigen}
G.~Guennebaud, B.~Jacob, {\em et~al.}, ``Eigen v3.''
  http://eigen.tuxfamily.org, 2010.

\bibitem{orava2012}
J.~Orava, A.~L. Greer, B.~Gholipour, D.~W. Hewak, and C.~E. Smith,
  ``Characterization of supercooled liquid {Ge2Sb2Te5} and its crystallization
  by ultrafast-heating calorimetry,'' {\em Nature Materials}, vol.~11,
  pp.~279--283, Apr. 2012.

\bibitem{reifenberg2006}
{J. Reifenberg}, {E. Pop}, {A. Gibby}, {S. Wong}, and {K. Goodson},
  ``Multiphysics {Modeling} and {Impact} of {Thermal} {Boundary} {Resistance}
  in {Phase} {Change} {Memory} {Devices},'' in {\em Thermal and
  {Thermomechanical} {Proceedings} 10th {Intersociety} {Conference} on
  {Phenomena} in {Electronics} {Systems}, 2006. {ITHERM} 2006.}, pp.~106--113,
  June 2006.

\bibitem{Pusztai05}
T.~Pusztai, G.~Bortel, and L.~Gr{\'a}n{\'a}sy, ``Phase field theory of
  polycrystalline solidification in three dimensions,'' {\em Europhys. Lett.},
  vol.~71, no.~1, pp.~131--137, 2005.

\bibitem{kheir2024}
O.~Abou El~Kheir, L.~Bonati, M.~Parrinello, and M.~Bernasconi, ``Unraveling the
  crystallization kinetics of the {Ge2Sb2Te5} phase change compound with a
  machine-learned interatomic potential,'' {\em npj Computational Materials},
  vol.~10, p.~33, Feb. 2024.

\bibitem{rahier2023}
E.~Rahier, M.-A. Luong, S.~Ran, N.~Ratel-Ramond, S.~Saha, C.~Mocuta, D.~Benoit,
  Y.~Le-Friec, and A.~Claverie, ``Multistep {Crystallization} of {Ge}-{Rich}
  {GST} {Unveiled} by {In} {Situ} synchrotron {X}-ray diffraction and
  (scanning) transmission electron microscopy,'' {\em physica status solidi
  (RRL) – Rapid Research Letters}, vol.~17, p.~2200450, Aug. 2023.

\bibitem{palumbo2017}
E.~Palumbo, P.~Zuliani, M.~Borghi, and R.~Annunziata, ``Forming operation in
  ge-rich gexsbytez phase change memories,'' {\em Solid-State Electronics},
  vol.~133, pp.~38--44, 2017.

\bibitem{sebastian2019}
A.~Sebastian, M.~Le~Gallo, and E.~Eleftheriou, ``Computational phase-change
  memory: beyond von {Neumann} computing,'' {\em Journal of Physics D: Applied
  Physics}, vol.~52, p.~443002, Aug. 2019.

\end{thebibliography}

\end{document}